\documentclass[reprint, amsmath, amssymb,aps,prb,floatfix, eqsecnum]{revtex4-2}

\usepackage[T1]{fontenc}
\usepackage{graphicx}
\usepackage{bm,array,amsmath,amsthm,booktabs}
\usepackage{amssymb}
\usepackage{hyperref}
\usepackage{color}
\usepackage{physics}
\usepackage[abs]{overpic}
\hypersetup{colorlinks = true,allcolors = blue }

\newcolumntype{L}{>{$}l<{$}}
\newcolumntype{C}{>{$}c<{$}}
\newcolumntype{R}{>{$}r<{$}}

\newcommand{\be}{\begin{eqnarray*}}
\newcommand{\ee}{\end{eqnarray*}}
\newcommand{\beq}{\begin{eqnarray}}
\newcommand{\eeq}{\end{eqnarray}}
\newcommand{\bequ}{\begin{equation}}
\newcommand{\eequ}{\end{equation}}
\newcommand{\bk}{{\mathbf{k}}}

\newcommand{\ph}{{\phantom{\dagger}}}

\newcommand{\zbb}{\mathbb{Z}}

\newcommand{\id}{\mathbb{I}}

\newcommand{\mct}{\mathcal{T}}
\newcommand{\mcc}{\mathcal{C}}
\newcommand{\mcs}{\mathcal{S}}

\DeclareMathOperator{\sgn}{sgn}

\newcommand{\tord}{T}
\newcommand{\antitord}{\widetilde{T}}

\newcommand{\pd}{\partial}
\newcommand{\bpmat}{\begin{pmatrix}}
\newcommand{\epmat}{\end{pmatrix}}
\newcommand{\bseq}{\begin{subequations}}
\newcommand{\eseq}{\end{subequations}}
\newcommand{\blp}{\boldsymbol{(}}
\newcommand{\brp}{\boldsymbol{)}}

\usepackage[english]{babel}

\usepackage{letltxmacro}

\LetLtxMacro{\ORIGselectlanguage}{\selectlanguage}
\makeatletter
\DeclareRobustCommand{\selectlanguage}[1]{%
  \@ifundefined{alias@\string#1}
    {\ORIGselectlanguage{#1}}
    {\begingroup\edef\x{\endgroup
       \noexpand\ORIGselectlanguage{\@nameuse{alias@#1}}}\x}%
}
\newcommand{\definelanguagealias}[2]{%
  \@namedef{alias@#1}{#2}%
}
\makeatother

\definelanguagealias{en}{english}

\newtheorem*{definition}{Definition}

\begin{document}
\title{Classification of unitary operators by local generatability}

\author{Xu Liu}
\thanks{These authors contributed equally to this work.}
\author{Adrian B. Culver{\color{blue}\scriptsize{*}}}
\email[Present address: Center for Quantum Science and Engineering, Department of Electrical and Computer Engineering, University of California, Los Angeles, Los Angeles, California 90095, USA,\qquad]{adrianculver@g.ucla.edu}
\author{Fenner Harper}
\author{Rahul Roy}
\email{rroy@physics.ucla.edu}
\affiliation{Mani L. Bhaumik Institute for Theoretical Physics, Department of Physics and Astronomy, University of California, Los Angeles, Los Angeles, California 90095, USA}

\date{\today}
\begin{abstract}
Periodically driven (Floquet) systems can exhibit possibilities beyond what can be obtained in equilibrium.  Both in Floquet systems and in the related problems of discrete-time quantum walks and quantum cellular automata, a basic distinction arises among unitary time evolution operators: while all physical operators are local, not all are locally generated (i.e., generated by some local Hamiltonian). In this paper, we define the notion of equivalence up to a locally generated unitary in all Altland-Zirnbauer symmetry classes.  We then classify noninteracting unitaries in all dimensions on this basis by showing that equivalence up to a locally generated unitary is identical to homotopy equivalence. 
\end{abstract}
\maketitle

\setcounter{footnote}{0} 
\footnotetext[0]{The classification in Ref.~\cite{GrossIndex2012} was obtained under the assumption of strict spatial locality, that is, unitaries strictly vanishing outside a finite interval.  The classification was later generalized in Ref.~\cite{CedzichChiral2021} to allow for exponential decay.}

\section{Introduction \label{sec:Intro}}

Gapped topological phases have been extensively studied, especially since the discovery of topological insulators, both because of the fundamental interest of finding phases that depart from the standard Landau paradigm and because of their potential for applications such as spintronics and quantum computing~\cite{HasanColloquium2010}. Defining characteristics of gapped topological phases include robustness to disorder, quantized transport signatures, and protected gapless surface states that cannot be realized in the bulk.  The classification of these phases proceeds by identifying homotopy equivalent (HE) Hamiltonians, where the homotopy must preserve the gap and any symmetries present.  In the noninteracting case, the classification is given by the periodic table of topological insulators and superconductors in the ten Altland-Zirnbauer (AZ) symmetry classes~\cite{ChiuClassification2016}, which can be obtained by algebraic K-theory (or other methods).  In the case of interacting topological phases protected by symmetry, group cohomology plays a key role in the classification~\cite{ChenLocal2010}.

A more recent focus has been the exploration of topological phases that become possible in nonequilibrium settings, especially in the paradigmatic case of a time-periodic Hamiltonian (Floquet) \cite{HarperTopology2020}.  Here the more natural object of study is not the Hamiltonian itself, but rather the unitary operator that implements time evolution.  Distinct Floquet topological phases correspond to homotopy-inequivalent unitary evolutions through the driving period, and are also related to the time evolution unitary for a full period (the Floquet operator) restricted to the edge.  Floquet topological phases have been realized experimentally in various platforms, such as ultracold matter~\cite{EckardtColloquium2017, JotzuExperimental2014} and photonic and acoustic systems~\cite{EsmannTopological2018,YangTopological2015,OzawaTopological2019}.

Floquet operators are local and, by definition, locally generated; that is, they are obtained from a local (and generally time-dependent) Hamiltonian.  However, many nonequilibrium settings feature unitaries that are local, but not locally generated.  Such operators have been studied in topological Floquet systems, where they can arise at the edge, and in the closely related problems of discrete-time quantum walks and quantum cellular automata (which correspond to the noninteracting and interacting case, respectively) \cite{GrossIndex2012,CedzichBulkedge2016,CedzichTopological2018,CiracMatrix2017,HigashikawaFloquet2019,GongClassification2020,Farrellyreview2020,CedzichChiral2021,HaahTopological2022}. In one dimension with no symmetries, a complete classification of local unitaries has been obtained, in both the noninteracting and interacting cases, by Gross \emph{et al.}~\cite{GrossIndex2012,Note0} (see also Ref.~\cite{KitaevAnyons2006} for earlier work on the noninteracting case).

An intriguing feature of the classification of Ref.~\cite{GrossIndex2012} is that it is the same regardless of which of the following notions of equivalence is used: Equivalence up to homotopy (note that unlike the Hamiltonian case there is no gap condition that the homotopy must preserve), or equivalence up to transformation by a locally generated unitary.  In the case of symmetries, however, the idea of equivalence up to transformation by a locally generated unitary (and its relation to homotopy equivalence) does not seem to have been discussed in the literature, even for noninteracting systems (although, see Ref.~\cite{GongClassification2020}).  (Let us note here that the notion of equivalence up to transformation by a locally generated unitary also appears in the classification of interacting topological phases in the static case \cite{ChenLocal2010}, and that locally generated unitaries are related to finite depth quantum circuits.) 

In this paper, we classify noninteracting, local unitaries up to equivalence by local generatability in all ten AZ symmetry classes and in all spatial dimensions.  A key point of our work is that we define the notion of equivalence of unitaries up to local generatability in the case of symmetries.  (We show in the main text that the naive extension of the existing definition in the case without symmetries does not work in most of the symmetry classes.)  Furthermore, we show that equivalence up to local generatability is identical to equivalence up to homotopy.  We then solve the homotopy classification problem using a mapping between unitaries and Hamiltonians that was introduced in the study of Floquet phases in Ref.~\cite{RoyPeriodic2017}.  We thus obtain a periodic table for local unitaries, providing the group of possible (strong) topological invariants in each dimension and each symmetry class, with a classification that can be regarded interchangeably as being based on equivalence up to local generatability or on homotopy equivalence. 

We note that in Ref.~\cite{HigashikawaFloquet2019}, Higashikawa \emph{et al.} obtained a homotopy classification of gapless bulk Floquet states, resulting in the same periodic table as we obtain here.  We comment in more detail on the relation of our work to that of Ref.~\cite{HigashikawaFloquet2019} in Sec. \ref{sec:top_class_unitaries}.  In the 1D interacting case, Ref.~\cite{GongClassification2020} has studied equivalance up to local generatabilty, allowing for unitary symmetries, in the context of matrix-product unitaries.  An alternative approach to nonequilibrium classification, based on wavefunction topology, was presented in Ref.~\cite{McGinleyClassification2019}.  We also note that some results in this paper were reported by two of the current authors in Ref.~\cite{LiuClassification2020}.

The paper is organized as follows.  In Sec. \ref{sec:Prelim}, we describe the setting of our work in more detail and provide the necessary background on the AZ symmetry classes.  In Sec. \ref{sec:Unitaries in class A: Local Generatability, Homotopy, and Classification}, we consider unitaries without symmetries as a first illustration of our approach, showing the equivalence of local generatability equivalence to homotopy equivalence and obtaining the  topological classification in all spatial dimensions.  We then present the same approach in the case of symmetries in two steps: In Sec. \ref{sec:Local generatability and homotopy for unitaries with symmetry}, we define the notion of equivalence up to local generatability in all symmetry classes and show that it is equivalent to homotopy equivalence; then, in Sec. \ref{sec:Classification of local unitaries}, we classify local unitaries up to homotopy.  In Sec. \ref{sec:examples}, we provide some example lattice models that yield topologically nontrivial unitaries.  We provide discussion and outlook in Sec. \ref{sec:conclusion}.

\section{Background\label{sec:Prelim}}

\subsection{Locality and local generatability\label{sec:local_operators_class_A}}
In this section, we introduce the definitions and preliminary ideas that are used in the rest of the paper.  The basic setting is an infinite real-space lattice of any dimension with sites labeled by $j,k,\dots$ and with a fixed number of generalized orbitals at each site (corresponding to, e.g., orbital and spin degrees of freedom).  Translation invariance is not assumed.  A unitary operator $U$ (also referred to as a ``unitary'') is defined by the usual condition
\beq
U^\dagger U =UU^\dagger&=&\id,
\eeq
where $\dagger$ indicates the Hermitian conjugate and $\id$ is the identity operator.

We define a \emph{local} unitary to be one whose matrix elements $U_{jk}$ (where generalized orbital quantum numbers are suppressed) decay exponentially or faster at large separation in real space. Explicitly, we require that
\beq
\left|U_{jk}\right|&\leq&Ce^{-\left|j-k\right|/\ell}\label{eq:locality}
\eeq
for some positive constants $C$ and $\ell$ and for large enough $\left|j-k\right|$, where $\left|j-k\right|$ is shorthand for the distance between the sites $j$ and $k$ and where $\ell$ may be regarded as a sort of localization length of $U$.  Let us also note that when we refer to Hamiltonians (i.e., Hermitian operators) being local, we mean that their real-space matrix elements decay exponentially as in Eq.~\eqref{eq:locality}.

The classification that we derive in this paper describes unitaries that are local, but not necessarily \emph{locally generated}. For our purposes, we define a locally generated unitary as one which may be written as the time evolution by some local Hamiltonian $H(t)$:
\beq
U_\text{LG}&=&\tord\exp\left[-i\int_0^1 dt\ H(t) \right],\label{eq:locally_generated}
\eeq
where $\tord$ is the time-ordering symbol (not to be confused with the time reversal symmetry operator $\mct$ discussed below) and where we set an arbitrary timescale to unity for convenience.  Throughout the paper, we use the shorthand expression ``$U_\text{LG}$ is generated by $H(t)$'' to refer to Eq.~\eqref{eq:locally_generated}.  We also use the expression ``$U_\text{LG}(s)$ is generated by $H(t)$ from $t=0$ to $s$'' (where $0\le s \le 1$) to refer to
\beq
U_\text{LG}(s)=\tord\exp\left[-i\int_0^sdt\  H(t)\right]\label{eq:locally_generated_from_0_to_s}.
\eeq
We include the subscript LG to distinguish the locally generated case from the more general case of a family of unitaries $U(s)$ that may or may not be locally generated.

Importantly, a locally generated unitary is always local due to a time-dependent, single-particle version of the Lieb-Robinson theorem~\cite{GrafBulk2018}, which can be simply stated as: The time propagator is local when the Hamiltonian is.  Although Ref.~\cite{GrafBulk2018} studies two-dimensional systems, the proof there does not require the time propagator to be two-dimensional and thus can be generalized to arbitrary dimensions. 

The concept of a locally generated unitary discussed here is closely related to the concept of a finite depth quantum circuit in the context of quantum information.  It is also related to the notion of local implementability discussed in Ref.~\cite{GrossIndex2012}. A time evolution unitary is called locally implementable in Ref.~\cite{GrossIndex2012} if it can be achieved by a finite number of commuting block unitaries (each of finite length), i.e., quantum gates, or by a finite product of partitioned operations.  Locally generated unitaries, as defined by Eq.~\eqref{eq:locally_generated}, encompass both finite depth quantum circuits and locally implementable time evolutions.

Let us note that our topological classification will not make explicit reference to the idea of repeating a single unitary many times as a discrete-time evolution; our focus throughout is on the unitary itself.  Thus, we need not consider the question of whether or not the $H(t)$, that appears in Eq.~\eqref{eq:locally_generated} extends periodically for $t$ beyond $1$.

\subsection{Symmetry operators and symmetry classes for gapped Hamiltonians\label{sec:HermitianSymmetries}}
The well-known periodic table of topological insulators and superconductors~\cite{KitaevPeriodic2009} is a classification of gapped, free-fermion Hamiltonians, arranged according to spatial dimension and symmetry class.  Let us recall that a \emph{gapped} Hamiltonian is one whose eigenvalue spectrum has a finite gap around zero.  Explicitly, if $\{E_i\}$ are the (necessarily real) eigenvalues of a Hamiltonian $H$, then $H$ is gapped if and only if
\beq
\left|E_i\right|>E_c
\eeq
for some positive $E_c$, for all indices $i$.

The symmetry classes included in the periodic table label the presence or absence of three nonspatial symmetries (see Ref.~\cite{ChiuClassification2016} for a review): Time reversal symmetry (TRS), particle-hole symmetry (PHS), and chiral symmetry (CS, also known as sublattice symmetry), with symmetry operators denoted as $\mct$, $\mcc$, and $\mcs$, respectively.  The first two of these symmetry operators are antiunitary (i.e., proportional to the complex conjugation operator), and, if present, act on a real-space Hamiltonian according to
\bseq
\begin{align}
\mct H\mct^{-1}&=H,\label{eq:TRS_ham}\\
\mcc H\mcc^{-1}&=-H.\label{eq:PHS_ham}
\end{align}
\eseq
CS is a unitary symmetry which is present automatically if both TRS and PHS are present:
\beq
    \mcs = e^{i\delta} \mcc\mct,\label{eq:CS_as_CT}  
\eeq
where the phase $\delta$ can be set by some convention for each symmetry class; even if TRS and PHS are not both present, CS may be present independently. CS has the action
\beq
\mcs H \mcs^{-1}&=&-H.\label{eq:CS_ham}
\eeq
The antiunitary symmetries $\mct$ and $\mcc$ may only square to either $+\id$ or $-\id$, which are physically distinct cases, while the unitary symmetry $\mcs$ squares to the identity up to an arbitrary phase factor. The presence or absence of each symmetry, along with the sign of the squares of $\mct$ and $\mcc$, yields the ten AZ symmetry classes \cite{AltlandNonstandard1997}. We note that unitary commuting symmetries are ignored in this classification; if present, they allow the Hamiltonian to be block diagonalized, and then each block can be treated separately.

For later convenience, we now fix our phase convention for CS.  We assume that the CS operator is traceless, as in Refs.~\cite{GrafBulk2018,KitaevPeriodic2009}.  Then, one can always bring the CS operator to the form $\mcs = e^{i\varphi}\sigma_z \otimes \id$ for some phase $\varphi$.  In the four symmetry classes with both TRS and PHS, consistency with Eq.~\eqref{eq:CS_as_CT} requires $e^{2i\varphi}=e^{2i\delta}(\sgn \mcc^2)(\sgn \mct^2)$ (where we have recalled that $\mct$ and $\mcc$ can always be made to commute with each other).  Throughout, we use the convention $\varphi=0$, which in turn requires $\delta$ to vary depending on the symmetry class; this allows us to use the same expression
\bequ
    \mcs = \sigma_z \otimes \id\label{eq:CS_canonical_form}
\eequ
in all symmetry classes (as in, e.g., Refs.~\cite{Katsuranoncommutative2018,ChungTopological2023}).  In the symmetry classes with $\mct^2\ne \mcc^2$, we then must have $e^{i\delta}=\pm i$, so $\mcs$ anticommutes with $\mct$ and with $\mcc$.  Note also that
\bequ
    \mcs^{-1}= \mcs.\label{eq:CS_inverse}
\eequ
We emphasize that our phase convention is purely one of convenience; we could instead have chosen $\varphi$ so that $\mcs$ always commutes with $\mct$ and $\mcc$ [and had $\mcs^{-1}= e^{-2i\varphi}\mcs$ instead of Eq.~\eqref{eq:CS_inverse}].

 \subsection{Equivalence classes of gapped Hamiltonians\label{sec:hermitian_equivalence}}
Before conducting the classification of unitaries, we review the topological classification of gapped Hamiltonians. As we later use a one-to-one correspondence between Hamiltonians and unitaries \cite{RoyPeriodic2017}, this review can provide us tools to understand equivalence classes of unitaries.
 
Recall that two gapped Hamiltonians $H_0$ and $H_1$ are defined to be homotopy equivalent within a given symmetry class if and only if there exists a path $H(s)$ of local Hamiltonians, with $0\le s\le 1$ and each $H(s)$ in the symmetry class, which connects $H_0$ and $H_1$ smoothly without ever closing the energy gap \cite{KitaevPeriodic2009}.  The periodic table classification is based on the more general notion of \emph{stable} homotopy equivalence (see below).

A \emph{flattened} Hamiltonian is defined as one whose eigenvalues satisfy
\beq
E_i&\in&\{-1,+1\}
\eeq
for all indices $i$.  In particular, we note that this is true if and only if $H^2=\id$.  Flattened Hamiltonians are useful for classification purposes because any local, gapped Hamiltonian can be brought to a flattened form without breaking the locality \cite{ProdanBulk2016,Katsuranoncommutative2018} (albeit at the expense of introducing longer-ranged hopping terms).  This is done by ``spectral flattening,'' in which one defines
\beq
H^\prime=1-2P,\label{eq:Hamiltonian_flatten}
\eeq
where $P$ is the projector onto the occupied states (of a given Hamiltonian $H$).  The equivalence class of a Hamiltonian depends only on its spectral flattening. 

Table~\ref{tab:periodic} can most easily be generated using the methods of topological K-theory when the underlying system has lattice translation invariance, so that the periodicity of the Brillouin zone may be used and one may analyze vector bundles or bundles of matrices in the Brillouin zone. However, the entries in the classification (which describe strong topological phases) hold even when this symmetry is broken (e.g., by disorder). A rough argument for why this is the case is that topological phases are labeled by discrete invariants that cannot vary continuously and are protected by a bulk energy gap. Adding weak disorder to a translation-invariant system acts as a small perturbation, which cannot change the value of this invariant unless it is strong enough to close the energy gap. The classification therefore naturally extends to systems with weak (symmetry-respecting) disorder. With a disorder that is strong enough that the energy gap closes, the topological phases can still survive, provided there remains a mobility gap \cite{Katsuranoncommutative2018}. In these situations, K-theoretical approaches can be applied, and real-space expressions for topological invariants may be used to diagnose the topology of a phase \cite{Katsuranoncommutative2018}.

For later use, let us define the stable homotopy relation.  The key point is that the homotopy relation ($\approx$) is only valid between Hamiltonians with the same number of generalized orbitals per site (i.e., the same number of bands in the translation invariant case).  Although, strictly speaking, the concept of stable homotopy is defined in K-theory only in the translation invariant case, we present the natural generalization.  The stable homotopy relation is denoted $\sim$ and defined as follows: $H_0\sim H_1$ if and only if there are trivial Hamiltonians $H_0^{(0)}$ and $H_1^{(0)}$ such that
\beq
H_0 \oplus H_0^{(0)} \approx H_1 \oplus H_1^{(0)}. \label{eq:stable_homotopy_Hamiltonians}
\eeq
Within any particular symmetry class, the trivial Hamiltonians here are required to be in that symmetry class.  Note that in the translation invariant case, the trivial Hamiltonians correspond to the addition of trivial bands.

\begin{table}
\be
\begin{array}{c|ccc|cccccccc}
\hline
\hline
\mbox{AZ class} & \mct & \mcc & \mcs & d=0 & 1 & 2 & 3 & 4 & 5 & 6 & 7\\
\hline\hline
\mbox{A} & 0 & 0 & 0 & \zbb & 0 & \zbb & 0 & \zbb & 0 & \zbb & 0\\
\mbox{AIII} & 0 & 0 & 1 & 0 & \zbb & 0 & \zbb & 0 & \zbb & 0 & \zbb\\
\hline
\mbox{AI} & + & 0 & 0 & \zbb & 0 & 0 & 0 & \zbb & 0 & \zbb_2 & \zbb_2 \\
\mbox{BDI} & + & + & 1 & \zbb_2 & \zbb & 0 & 0 & 0 & \zbb & 0 & \zbb_2 \\
\mbox{D} & 0 & + & 0 & \zbb_2 & \zbb_2 & \zbb & 0 & 0 & 0 & \zbb & 0 \\
\mbox{DIII} & - & + & 1 & 0 & \zbb_2 & \zbb_2 & \zbb & 0 & 0 & 0 & \zbb \\
\mbox{AII} & - & 0 & 0 & \zbb & 0 & \zbb_2 & \zbb_2 & \zbb & 0 & 0 & 0 \\
\mbox{CII} & - & - & 1 & 0 & \zbb & 0 & \zbb_2 & \zbb_2 & \zbb & 0 & 0 \\
\mbox{C} & 0 & - & 0 & 0 & 0 & \zbb & 0 & \zbb_2 & \zbb_2 & \zbb & 0 \\
\mbox{CI} & + & - & 1 & 0 & 0 & 0 & \zbb & 0 & \zbb_2 & \zbb_2 & \zbb \\
\hline\hline
\end{array}
\ee
\caption{Periodic table of topological insulators and superconductors \cite{KitaevPeriodic2009}. The leftmost column gives the letter label for each AZ symmetry class. The next three columns indicate the presence or absence of TRS ($\mct$), PHS ($\mcc$), and CS ($\mcs$) for each symmetry class, along with whether they square to $+\id$ or $-\id$. The rightmost eight columns indicate the topological classification for each symmetry class in dimension $d$. Note that the classification depends only on $d\mod{8}$ ($d \mod{2}$ for classes A and AIII) due to Bott periodicity \cite{KitaevPeriodic2009}. See main text for details.\label{tab:periodic}}
\end{table}

Let us comment here on the question of rigor.  Table~\ref{tab:periodic} has been obtained rigorously under the assumption of a spectral gap, which corresponds to weak disorder (see Ref.~\cite{ChungTopological2023} and references therein).  Under the assumption of a mobility gap, which corresponds to strong disorder, Ref.~\cite{ChungTopological2023} presents some progress toward a rigorous classification.  Although rigor is not our primary concern, we mention these points because an essential part of our paper is the mapping of the classification problem for unitaries to the classification problem for Hamiltonians.  Further improvement in the rigorous classification of Hamiltonians then directly translates to the classification of  unitaries.

\subsection{Symmetry operators and symmetry classes for unitaries \label{sec:unitary_symmetry}}
In this section, we define how the basic symmetries act on unitaries.  We do this by generalizing from the special case of locally generated unitaries.  In this special case, the actions of the basic symmetries on a unitary $U$ can be obtained from the actions of the symmetries on the time-dependent Hamiltonian $H(t)$ that generates $U$ \cite{CedzichChiral2021}.

We consider a unitary $U$ that is locally generated.  In particular, we consider $U= U_\text{LG}(1)$, where $U_\text{LG}(s)$ is generated by $H(t)$ from $t=0$ to $s$ [recall Eq.~\eqref{eq:locally_generated_from_0_to_s}].  Symmetries, if present, act on the instantaneous Hamiltonian as \cite{RoyPeriodic2017}
\bseq
\begin{align}
\mct H(t)\mct^{-1}&= H(1-t),\label{eq:TRS_Hamiltonian_floquet}\\
\mcc H(t)\mcc^{-1}&=-H(t),\label{eq:PHS_Hamiltonian_floquet}\\
\mcs H(t)\mcs^{-1}&=-H(1-t),\label{eq:CS_Hamiltonian_floquet}
\end{align}
\eseq
and therefore on the time evolution operator $U_\text{LG}(t)$ as \cite{RoyPeriodic2017}
\bseq
\begin{align}
\mct U_\text{LG}(t)\mct^{-1}&=U_\text{LG}(1-t)U_\text{LG}^\dagger(1),\label{eq:TRS_unitary_floquet}\\
\mcc U_\text{LG}(t)\mcc^{-1}&=U_\text{LG}(t),\label{eq:PHS_unitary_floquet}\\
\mcs U_\text{LG}(t)\mcs^{-1}&=U_\text{LG}(1-t)U_\text{LG}^\dagger(1).\label{eq:CS_unitary_floquet}
\end{align}
\eseq
Setting $t=1$ and noting that $U_\text{LG}(0)=\mathbb{I}$, we obtain
\bseq
\begin{align}
\mct U\mct^{-1}&=U^\dagger,\label{eq:TRS_unitary}\\
\mcc U\mcc^{-1}&=U,\label{eq:PHS_unitary}\\
\mcs U\mcs^{-1}&=U^\dagger.\label{eq:CS_unitary}
\end{align}
\eseq

Generalizing from this special case, we then \emph{define} the actions of the three symmetries (if present) on a unitary $U$ to be given by \eqref{eq:TRS_unitary}--\eqref{eq:CS_unitary}, whether or not $U$ is locally generated.  With these symmetry definitions, we can assign unitaries to the same ten AZ symmetry classes as static Hamiltonians.

The action of CS has been obtained this way in Ref.~\cite{CedzichChiral2021}.  Also, the actions of all symmetries have been obtained in Refs.~\cite{CedzichBulkedge2016,CedzichTopological2018} by considering the symmetry relations for static Hamiltonians [$H(t) = H$], as we now review.  It is straightforward to show that any symmetries of $H$ [Eqs.~\eqref{eq:TRS_ham}, \eqref{eq:PHS_ham}, and \eqref{eq:CS_ham}] yield the following symmetry properties of $e^{-i H}$:
\bseq
\begin{align}
\mct e^{-iH} \mct^{-1}&=e^{+iH},\\
\mcc e^{-iH} \mcc^{-1}&=e^{-iH},\\
\mcs e^{-iH} \mcs^{-1} &=e^{+iH},
\end{align}
\eseq
which indeed agree with Eqs.~\eqref{eq:TRS_unitary}--\eqref{eq:CS_unitary} in the special case $U= e^{-i H}$.

\section{Unitaries in class A: Local Generatability, Homotopy, and Classification\label{sec:Unitaries in class A: Local Generatability, Homotopy, and Classification}}

As a warm-up to the case of unitaries with symmetry, we consider unitaries without symmetry (class A).  We define two notions of equivalence: one based on local generatability and the other based on homotopy.  Our goal is to classify unitaries according to equivalence up to local generatability.  To achieve this, we first prove that two unitaries in class A are equivalent up to local generatability if and only if they are homotopy equivalent.  Thus, it suffices to classify unitaries up to homotopy equivalence.

We then construct a one-to-one mapping between unitaries in class A and flattened Hamiltonians in class AIII~\cite{RoyPeriodic2017}, and we show that this mapping preserves homotopy equivalence (that is, two unitaries are homotopy equivalent if and only if the corresponding two Hamiltonians are homotopy equivalent).  We can then read off the classification for class A unitaries in all dimensions from the existing classification of class AIII Hamiltonians.  (We are using the well-known fact that flattened Hamiltonians follow the same classification as the larger set of gapped Hamiltonians.)  In sum, the homotopy definition of equivalence is more convenient for importing known results for Hamiltonians, and the definition of equivalence up to local generatability provides an alternate, equivalent interpretation of the resulting classification of unitaries.  In this way, we will generalize the classification of Ref.~\cite{GrossIndex2012} from the one-dimensional case to all spatial dimensions (in the noninteracting case). 

\subsection{Local generatability and homotopy in class A\label{sec:Local generatability and homotopy in class A}}

We define $U_0$ and $U_1$ to be locally generated equivalent (LGE) if and only if $U_1= U_{\text{LG}}U_0 $ for some $U_{\text{LG}}$ that is locally generated [recall Eq.~\eqref{eq:locally_generated}].  Note that, according to this definition, a unitary is locally generated if and only if it is LGE to the identity.

We define two unitaries $U_0$ and $U_1$ to be HE if and only if there is a homotopy between them, i.e., there is a family of unitaries $U(s)$ ($0\le s \le 1$) with $U(0)=U_0$, $U(1)=U_1$ and with sufficiently nice dependence on $s$.  We do not attempt to find a precise definition of sufficiently nice, but will assume at least that $U(s)$ is piecewise-differentiable.

Homotopy equivalence is manifestly an equivalence relation, i.e., it is reflexive, symmetric, and transitive.  Although these properties can also be verified directly for locally-generated equivalence, it is simpler to note that they follow from our demonstration below.

We proceed to show that LGE and HE are equivalent in class A.  We start with the more straightforward direction of the proof, which is to show that two unitaries $U_0$ and $U_1$ that are LGE must also be HE.  By assumption, $U_1= U_\text{LG}U_0$, where $U_\text{LG}$ is generated by some Hamiltonian $H(t)$.  We define $U_\text{LG}(s)$ to be generated by $H(t)$ from $t=0$ to $s$; then we define a homotopy from $U_0$ to $U_1$ by $U(s)= U_\text{LG}(s)U_0$.  Thus, $U_0$ and $U_1$ are HE.

Suppose instead that $U_0$ and $U_1$ are HE.  We then define
\bequ
    H(s) = i \frac{\partial U(s)}{\partial s}U^\dagger(s),\label{eq:H(s)_simple_def}
\eequ
where $U(s)$ is the given homotopy from $U_0$ to $U_1$.  To see that $H(s)$ is a local Hamiltonian, we note that it is Hermitian due to $U(s)$ being unitary and that it is local because $U(s)$ is and because of the assumption that the homotopy is sufficiently nice in $s$  [which we understand to imply that $\pd_s U(s)$ inherits locality from $U(s)$].

By construction, $U(s)$ satisfies
\bequ
    i \frac{\pd U(s)}{\pd s} = H(s)U(s).\label{eq:diff_eqn_A}
\eequ
The key point is that \eqref{eq:diff_eqn_A} is a first-order differential equation and hence has a unique solution given the initial condition $U(0)= U_0$.  The solution is
\beq
    U(s) = \tord \exp[-i \int_0^s dt\ H(t) ]U_0.\label{eq:ansatz_A}
\eeq
Setting $s=1$, we see that $U_1= U_\text{LG}U_0$, where $U_\text{LG}$ is generated by $H(t)$; this confirms that $U_0$ and $U_1$ are LGE.  Thus, we have shown that LGE and HE are equivalent definitions for an equivalence relation, denoted $\approx$, on unitaries in class A.

The equivalence relation $\approx$ can only relate unitaries with the same number of generalized orbitals per lattice site.  We now define a more general notion of \emph{stable} equivalence (denoted $\sim$) that can relate unitaries with different numbers of generalized orbitals per lattice site.  This more general notion is necessary for making contact with the stable homotopy equivalence relation that is used in the classification of Hamiltonians [see Eq.~\eqref{eq:stable_homotopy_Hamiltonians} and the discussion there].

We define $U_0 \sim U_1$ if and only if there exist two trivial (i.e., locally generated) unitaries $U_0^{(0)}$ and $U_1^{(0)}$ such that 
\beq
U_0 \oplus U_0^{(0)} \approx U_1 \oplus U_1^{(0)},\label{eq:def_stable_equivalence_of_unitaries}
\eeq
where $\oplus$ is the direct sum.  (A related definition was provided in Ref.~\cite{RoyPeriodic2017} for translation-invariant unitary evolutions satisfying a gap condition.)  We thus have two equivalent definitions of the equivalence relation $\sim$: ``stable LGE'' and ``stable HE'' (corresponding to $\approx$ being understood as LGE or HE).

We emphasize that the above definition of equivalence will lead to a different classification from that which arises in the study of Floquet systems (such as Ref.~\cite{RoyPeriodic2017}). The topological classification of Floquet systems informs us about the presence of protected edge modes in gaps in the quasienergy spectrum. However, since these unitary Floquet operators are obtained by evolving with a local Hamiltonian, they are trivial according to our definition. Instead, the classification we study in this paper will lead to unitaries that are, in general, gapless and not locally generated.  In this way, the nontrivial unitaries that we discuss cannot arise in the bulk of a Floquet time evolution, but could arise when the Floquet operator is restricted to the \emph{boundary} of a time-evolved physical system.

\subsection{Classification in class A\label{sec:Classification in class A}}
Our goal is to classify unitaries by LGE.  To do this, we use a mapping between unitaries and flattened Hamiltonians~\cite{RoyPeriodic2017}.  Given a unitary $U$ in class A, we define a Hamiltonian $H_U$ that acts on two copies of the Hilbert space of $U$ as follows~\cite{RoyPeriodic2017}: 
\beq
H_U&=&\left(\begin{array}{cc}
0 & U\\
U^\dagger & 0
\end{array}\right).\label{eq:H_U_def}
\eeq
We can interpret this Hilbert space doubling as the addition of a sublattice degree of freedom to each site, which we label $A$ and $B$.  By construction, $H_U$ is a Hamiltonian ($H_U^\dagger= H_U$), and $H_U$ is flattened because $U$ is unitary.  Furthermore, $H_U$ is a local operator, inheriting its locality from $U$. Specifically, since the matrix elements $U_{ij}$ satisfy the locality condition in Eq.~\eqref{eq:locality}, so do the matrix elements $\left[H_U\right]_{\alpha i,\beta j}$, where we have included the sublattice indices $\alpha,\beta\in\{A,B\}$ (while suppressing any other generalized orbital indices common to both $U$ and $H_U$).

The Hamiltonian $H_U$ automatically has CS, with the symmetry operator given by
\beq
\mcs'&=&\left(\begin{array}{cc}
\id & 0\\
0 & -\id
\end{array}\right),\label{eq:mcsprime_classA}
\eeq
i.e., $\sigma_z$ in the sublattice basis.  To verify that $\mcs'$ is indeed a CS operator, we note that it is unitary, squares to $\id$, and has the action
\beq
\mcs'H_U\left(\mcs'\right)^{-1}=\left(\begin{array}{cc}
0 & -U\\
-U^\dagger & 0
\end{array}\right)=-H_U,\label{eq:CS_H_U}
\eeq
in agreement with Eq.~\eqref{eq:CS_ham}.  Thus, the mapping $U\mapsto H_U$ sends class A unitaries to flattened, class AIII Hamiltonians.  (See the first row of Table~\ref{tab:symmetry_transformations}.)

Next we show that the mapping is one-to-one by constructing the inverse.  Without loss of generality, the CS operator in class AIII may be taken to be Eq.~\eqref{eq:mcsprime_classA}.  It is then straightforward to show that a flattened Hamiltonian satisfying Eq.~\eqref{eq:CS_H_U} \emph{must} take the form of $H_U$ as given in Eq.~\eqref{eq:H_U_def} with $U$ being unitary (see Appendix \ref{app:Further details on the mapping from non-chirally-symmetric unitaries to Hamiltonians}).  The locality of $U$ follows from the locality of $H_U$.  Thus, we have defined a mapping $H_U\to U$ which is the inverse of \eqref{eq:H_U_def}.

Importantly, the mapping \eqref{eq:H_U_def} preserves stable homotopy equivalence.  That is,
\beq
    U_0 \sim U_1 \text{ if and only if } H_{U_0} \sim H_{U_1}.\label{eq:mapping preserves stable homotopy}
\eeq
This equivalence can be read off essentially by inspection of Eq.~\eqref{eq:H_U_def}.  We present a more detailed verification in Appendix \ref{app:Mapping between unitaries and Hamiltonians preserves stable homotopy} (where we in fact treat the case of a general symmetry class). 

By \eqref{eq:mapping preserves stable homotopy}, we can read off the classification of unitaries in class A from the existing classification of Hamiltonians in class AIII.  From the second row of Table ~\ref{tab:periodic}, we read off that class A unitaries have a $\mathbb{Z}$ classification in all odd-numbered spatial dimensions of strong topological phases and a trivial classification in all even-numbered spatial dimensions (yielding the first row in Table \ref{tab:periodic_unitary} below).  As a check, we note that a $\mathbb{Z}$ classification was obtained in the one-dimensional case in Refs.~\cite{KitaevAnyons2006,GrossIndex2012}, and a trivial classification in the zero-dimensional case in Ref.~\cite{CedzichTopological2018}.  (In the latter case, we refer to the part of Ref.~\cite{CedzichTopological2018} that considers one-dimensional unitaries satisfying a gap condition; this condition effectively makes these unitaries correspond to zero-dimensional unitaries in our case.  Table \ref{tab:periodic_unitary} below confirms this correspondence in all other symmetry classes, as well.)

Thus, we have shown that unitaries without symmetry can be classified by equivalence up to local generatability (i.e., LGE).  We obtained this classification in two steps: (1) we showed the equivalence of LGE to HE; and (2) we used a one-to-one mapping from nonsymmetric unitaries to chirally-symmetric, flattened Hamiltonians to read off (from known results for Hamiltonians) the classification of nonsymmetric unitaries by HE, which by (1) is equivalent to the classification by LGE.

Our task in the next two sections is to repeat these steps, with suitable adjustments to account for symmetries, for unitaries in the remaining nine symmetry classes.  

\section{Local generatability and homotopy for unitaries with symmetry}\label{sec:Local generatability and homotopy for unitaries with symmetry}

When are two unitaries in the same symmetry class topologically equivalent?  As in the class A case discussed in Sec. \ref{sec:Unitaries in class A: Local Generatability, Homotopy, and Classification}, we present two definitions (LGE and HE) that turn out to be equivalent.  As we discuss below, the definitions of HE in general and of LGE \emph{to the identity} are, in all symmetry classes, obvious generalizations of the class A case.  However, in most of the symmetry classes it is not obvious how to define LGE between two arbitrary unitaries, and indeed one of our main results is defining LGE throughout all symmetry classes and showing that it is equivalent to HE. 

\subsection{LGE and HE for unitaries with symmetry}
We consider an arbitrary AZ symmetry class $\mathbb{S}$.
\begin{definition}
Two unitaries $U_0$ and $U_1$ are HE in $\mathbb{S}$ if and only if there is a homotopy between them within $\mathbb{S}$, i.e., there is a family of unitaries $U(s)$ ($0\le s \le 1$) with $U(0)=U_0$, $U(1)=U_1$, sufficiently nice dependence on $s$, and each $U(s)$ in $\mathbb{S}$.
\end{definition}

In the case $\mathbb{S}=$AIII, for example, each $U(s)$ is required to satisfy $\mcs U(s) \mcs^{-1}=U^\dagger(s)$ [c.f. Eq.~\eqref{eq:CS_unitary}].

To define the equivalence of two unitaries up to local generatability, we first define a unitary $U$ to be locally generated in class $\mathbb{S}$ if and only if $U$ is generated by $H(t)$ for some $H(t)$ in $\mathbb{S}$.  [For instance, if $\mathbb{S}=$AIII, then $H(t)$ must satisfy Eq.~\eqref{eq:CS_Hamiltonian_floquet}.]  Both this definition and the definition of HE given in the preceding paragraph are clear generalizations from the class A case.

Next we define equivalence up to local generatability in class $\mathbb{S}$.  Before doing this in the general case, let us consider the two symmetry classes with only PHS (classes C and D).  In these cases, we proceed similarly as in the class A case: we define two unitaries $U_0$ and $U_1$ in class $\mathbb{S}=$C,D to be LGE if and only if $U_1=  U_{\text{LG}} U_0$ for some $U_{\text{LG}}$ that is locally generated in $\mathbb{S}$.  For this definition to be consistent, the product $U_\text{LG} U_0$ must satisfy PHS [Eq.~\eqref{eq:PHS_unitary}]; this follows immediately from the PHS of $U_0$ and of $U_\text{LG}$. 

The above definition does not work, however, in the symmetry classes with CS, TRS, or both.  The problem is that even if $U_0$ and $U_\text{LG}$ have (e.g.) CS, the product $U_\text{LG}U_0$ need not: $SU_\text{LG}U_0 S^{-1}=  U_\text{LG}^\dagger U_0^\dagger$ is not generally equal to $(U_\text{LG}U_0)^\dagger=U_0^\dagger U_\text{LG}^\dagger$ (and the case of TRS is similar).  The essential step in defining LGE is to replace the product of $U_0$ and $U_\text{LG}$ by a symmetry-preserving operation called \emph{composition}.  We provide the definition for this operation below, noting that it is motivated by a composition operation defined for Floquet drives in Ref.~\cite{RoyPeriodic2017}.  In a similar way as we motivated Eqs.~\eqref{eq:TRS_unitary}--\eqref{eq:CS_unitary} by generalizing from the case of a locally generated unitary, we can motivate the following definition as generalizing the composition operation from the case of two unitaries that are both locally generated (which is essentially covered by Ref.~\cite{RoyPeriodic2017}) to the case of one unitary that is locally generated and one that may or may not be.

For a class $\mathbb{S}$ with CS, TRS, or both, we define the composition of $U$ (in $\mathbb{S}$) with $U_\text{LG}$ (locally generated in $\mathbb{S}$) to be
\begin{widetext}
\beq
    U \star U_\text{LG} = \left(\tord e^{-i\int_{1/2}^1 dt\ H(t) } \right) U \left( \tord e^{-i\int_0^{1/2}dt\ H(t) }\right),\label{eq:composition_with_LG}
\eeq
\end{widetext}
where $H(t)$ generates $U_\text{LG}$.  It is straightforward to verify that $U \star U_\text{LG}$ as defined here is in $\mathbb{S}$ (see Appendix \ref{app:Composition preserves symmetry}); that is, the composition operation preserves symmetry.

Strictly speaking, $U \star U_\text{LG}$ depends on the choice of drive $H(t)$, in $\mathbb{S}$, that generates $U_{\text{LG}}$; this choice is generally nonunique.  In our calculations below, we can indeed regard the composition operation as having a unitary $U$ and a drive $H(t)$ as its inputs (i.e., we need not discuss the unitary $U_\text{LG}$ generated by the drive).  We nonetheless use the $U_\text{LG}$ notation because of the following simple result (shown in Appendix \ref{app:Composition is unique up to homotopy}): For any two choices of drives in $\mathbb{S}$ that generate a given $U_\text{LG}$ in $\mathbb{S}$, the corresponding two unitaries defined by the right-hand side of \eqref{eq:composition_with_LG} are HE in $\mathbb{S}$.  In other words, given $U$ and $U_\text{LG}$, the right-hand side of Eq.~\eqref{eq:composition_with_LG} is unique up to homotopy.

In the classes with no symmetry or PHS only ($\mathbb{S}=$A, C, or D), we use a simpler definition of composition, consisting of the product of the two unitaries:
\beq
    U_0\star U_\text{LG} =  U_\text{LG}U_0 \qquad (\mathbb{S} \text{ without CS or TRS}).\label{eq:LGE_def_simpler}
\eeq
We have already noted that the product preserves PHS, if present.  Let us note that an alternate approach is also possible; we can use the same definition \eqref{eq:composition_with_LG} in \emph{all} symmetry classes.  The classification is the same in either case (see Appendix \ref{app:Classes without CS or TRS}).

Having defined a symmetry-preserving composition operation, we proceed to define equivalence up to local generatability in a general symmetry class $\mathbb{S}$.  This definition is one of the main points of our paper.
\begin{definition}
Two unitaries $U_0$
 and $U_1$ in $\mathbb{S}$ are LGE if and only if
 \beq
 U_1 = U_0\star U_\text{LG}\label{eq:LGE_def}
 \eeq
 for some $U_\text{LG}$ that is locally generated in $\mathbb{S}$.
\end{definition}

Let us make a few comments on this definition.  First, we emphasize that $U_0$ and $U_1$ here do not need to be locally generated.  Second, in the symmetry classes with CS, TRS, or both, it is understood to be sufficient for there to be \emph{some} choice of drive in $\mathbb{S}$ (that generates $U_{\text{LG}}$) for which Eq.~\eqref{eq:LGE_def} holds.  Third, in the special case $U_0=\mathbb{I}$, we have $U_0\star U_\text{LG}= U_\text{LG}$; thus, the unitaries that are LGE to the identity in $\mathbb{S}$ are precisely the unitaries that are locally generated in $\mathbb{S}$.  Finally, just as in the class A case, it is not necessary to verify that Eq.~\eqref{eq:LGE_def} indeed defines an equivalence relation (although this is straightforward); this follows from our demonstration below, since homotopy equivalence within $\mathbb{S}$ is clearly an equivalence relation.

\subsection{Proof that LGE and HE are equivalent}\label{sec:Proof that LGE and HE are equivalent}
We show that two unitaries $U_0$ and $U_1$ are LGE in $\mathbb{S}$ if and only if they are HE in $\mathbb{S}$.

We have already shown this equivalence in the case of  class A in Sec.~\ref{sec:Unitaries in class A: Local Generatability, Homotopy, and Classification}. For the two symmetry classes with PHS only ($\mathbb{S}=$C or D), the proof is very similar to the class A case.  Indeed, the only additional step is to verify that the Hamiltonian defined by Eq.~\eqref{eq:H(s)_simple_def} satisfies PHS [Eq.~\eqref{eq:PHS_Hamiltonian_floquet}]; this follows immediately from the antiunitarity of $\mcc$ and the PHS of $U(s)$.

We prove the equivalence of LGE to HE in the remaining seven symmetry classes in two overlapping groups: those that include CS (AIII, BDI, DIII, CII, and CI) and those that include TRS (AI, AII, BDI, DIII, CII, and CI).  For the four classes with both CS and TRS, we thus provide two proofs of the same result; we allow this redundancy because each proof contains an intermediate result that could be of interest.

Both proofs have the same structure: we show that each LGE and HE are equivalent to another condition.  To state this other condition, it is convenient to define $X$ to be the symmetry class obtained by removing all symmetries except PHS (if it is present) from a given class $\mathbb{S}$ that has CS, TRS, or both:
\beq
    X = 
    \begin{cases}
        \text{A} & \mathbb{S}= \text{AI, AII, or AIII},\\
        \text{D} & \mathbb{S} = \text{BDI or DIII},\\
        \text{C} & \mathbb{S} = \text{CI or CII}.
    \end{cases}\label{eq:X}
\eeq

Let us first consider the symmetry classes $\mathbb{S}$ that include CS. We claim that $U_0$ and $U_1$ are LGE if and only if there is some $U_\text{LG}^{(X)}$ that is locally generated in class $X$ for which
\beq
    U_1 = \mcs U_\text{LG}^{(X)\dagger}\mcs^{-1}U_0 U_\text{LG}^{(X)}.\label{eq:LG_canonical_relation_chiral}
\eeq
We further claim that $U_0$ and $U_1$ are HE if and only if the same condition \eqref{eq:LG_canonical_relation_chiral} holds.  Once we establish these two claims, it follows that LGE and HE are equivalent.

To motivate the condition \eqref{eq:LG_canonical_relation_chiral}, we consider here the special case of a unitary $U_1$ that is locally generated in a class with CS.  Equivalently, $U_1$ is LGE to the identity.  Our task is to obtain Eq.~\eqref{eq:LG_canonical_relation_chiral} with $U_0=\mathbb{I}$, providing some motivation for the case of general $U_0$.  To do this, we regard $U_1$ as the endpoint of a drive: $U_1= U_\text{LG}(1)$, where $U_\text{LG}(s)$ is generated from $t=0$ to $s$ by some $H(t)$ in the same symmetry class.  Specializing the CS property of the drive [Eq.~\eqref{eq:CS_unitary_floquet}] to $t=1/2$ and then rearranging yields $U_1= S U_\text{LG}^\dagger(\frac{1}{2})S^{-1}U_\text{LG}(\frac{1}{2})$, which is indeed Eq.~\eqref{eq:LG_canonical_relation_chiral} with $U_0=\mathbb{I}$ and $U_\text{LG}^{(X)}= U_\text{LG}(\frac{1}{2})$.  [An equivalent expression has been studied in Refs.~\cite{LiuChiral2018,CedzichChiral2021}.]  To confirm that $U_\text{LG}\left(\frac{1}{2}\right)$ is locally generated in $X$, we rescale $t\to t/2$ to show that $U_\text{LG}\left(\frac{1}{2}\right)$ is generated by
\beq
    H^{(X)}(t) \equiv \frac{1}{2}H(t/2),\label{eq:HX}
\eeq
which inherits PHS from $H(t)$ (if present).  Note that $H^{(X)}(t)$ cannot inherit CS or TRS from $H(t)$ because, as $t$ ranges from $0$ to $1$, $H^{(X)}(t)$ only includes the ``first half'' of $H(t)$.

By similar manipulations, we prove the more general statement that LGE is equivalent to \eqref{eq:LG_canonical_relation_chiral} in the symmetry classes that include CS.  The proof is similar to the previous paragraph and is presented in Appendix \ref{app:Classes with CS}.

It remains to show the equivalence of HE to the condition \eqref{eq:LG_canonical_relation_chiral}.  One direction is trivial: Given \eqref{eq:LG_canonical_relation_chiral}, the required homotopy may be defined using a homotopy $U_\text{LG}^{(X)}(s)$ that connects the identity to $U_\text{LG}^{(X)}$.  In particular, we use the class $X$ Hamiltonian $H^{(X)}(t)$ that generates $U_\text{LG}^{(X)}$ to define $U_\text{LG}^{(X)}(s)$ as being generated by $H^{(X)}(t)$ from $t=0$ to $s$.  Then $U_\text{LG}^{(X)}(s)$ is a family of unitaries that are locally generated in class $X$ with $U_\text{LG}^{(X)}(0)=\mathbb{I}$ and $U_\text{LG}^{(X)}(1)= U_\text{LG}^{(X)}$. We then construct the required homotopy from $U_0$ to $U_1$:
\beq
    U(s) = \mcs U_\text{LG}^{(X)\dagger}(s)\mcs^{-1} U_0  U_\text{LG}^{(X)}(s).\label{eq:homotopy_for_alternate_implies_HE}
\eeq
We may readily verify that $U(s)$ has PHS if present.  Also, $U(s)$ has CS [the proof is very similar to \eqref{eq:composition_preserves_CS} in the Appendix].  Thus, $U(s)$ is in the given symmetry class $\mathbb{S}$.

Next, we do the most substantial part of the proof: We show that HE (in a class with CS) implies \eqref{eq:LG_canonical_relation_chiral}.  Given a homotopy $U(s)$ that connects $U_0$ to $U_1$, we define
\beq
    h^{(X)}(s)= i\frac{1}{2} \frac{\pd U(s)}{\pd    s}U^\dagger(s).\label{eq:HX(s)}
\eeq
To verify that $h^{(X)}(s)$ is in class $X$, we note that $h^{(X)}(s)$ inherits PHS (if present) but cannot inherit CS or TRS [recall from Eqs.~\eqref{eq:TRS_unitary} and \eqref{eq:CS_unitary} that these latter two symmetries of $U(s)$ do not connect $s$ to $1-s$].  The locality of $h^{(X)}(s)$ follows from the locality of $U(s)$ in the same way as discussed below Eq.~\eqref{eq:H(s)_simple_def}.

By construction, we have $i \pd_s U(s)= 2i h^{(X)}(s) U(s)$.  We also have $i \pd_s U(s) = -2U(s)\mcs h^{(X)}(s) \mcs^{-1}$; to see this, we note that $\mcs h^{(X)}(s) \mcs^{-1}= i\frac{1}{2} [\pd_s U^\dagger(s)] U(s)= -i\frac{1}{2} U^\dagger(s) \pd_s U(s)$, where the first equality holds by the CS of $U(s)$ (and the homotopy being sufficiently nice to allow commuting $S$ with the derivative) and the second equality holds because $U(s)$ is unitary. Adding these equations yields
\beq
    i \frac{\pd U(s)}{\pd s} =   h^{(X)}(s)U(s) - U(s)\mcs h^{(X)}(s)\mcs^{-1}.\label{eq:diff_eqn_chiral}
\eeq
The key point is that Eq.~\eqref{eq:diff_eqn_chiral}, being a first-order differential equation, has a unique solution given the initial condition $U(0)= U_0$.  We can therefore obtain a convenient formal solution by making an ansatz and checking that it works.

Our ansatz is
\beq
    U(s) = u_\text{LG}^{(X)}(s) U_0 \mcs u_\text{LG}^{(X)\dagger}(s) \mcs^{-1},\label{eq:ansatz_chiral}
\eeq
where $u_\text{LG}^{(X)}(s)$ is generated by $h^{(X)}(t)$ from $t=0$ to $s$.  Then $h_\text{LG}^{(X)}(0)=\mathbb{I}$, so $U(0)= U_0$, as required.  The differential equation \eqref{eq:diff_eqn_chiral} holds due to $i\pd_s u_\text{LG}^{(X)}(s)= h^{(X)}(s)u_\text{LG}^{(X)}(s)$ (and the adjoint of this equation).  Thus, Eq.~\eqref{eq:ansatz_chiral} is a valid expression for the given homotopy $U(s)$.

We have therefore shown
\beq
    U_1 = u_\text{LG}^{(X)} U_0 \mcs u_\text{LG}^{(X)\dagger}\mcs^{-1},
\eeq
where $u_\text{LG}^{(X)}\equiv u_\text{LG}^{(X)}(1)$ is locally generated in class $X$.  As we now show, this form for $U_1$ is the same as \eqref{eq:LG_canonical_relation_chiral} up to trivial transformations.  We define $U_\text{LG}^{'(X)} = \mcs u_\text{LG}^{(X)\dagger}\mcs^{-1}$; then $U_1 = \mcs^{-1} U_\text{LG}^{'(X)\dagger} \mcs U_0  U_\text{LG}^{'(X)}$, which establishes Eq.~\eqref{eq:LG_canonical_relation_chiral} [c.f. Eq.~\eqref{eq:CS_inverse}].  Note that $U_\text{LG}^{'(X)}$ is generated by $-\mcs h^{(X)}(1-t) \mcs^{-1}$ and is thus locally generated in class $X$.  This completes the proof of the equivalence of HE to \eqref{eq:LG_canonical_relation_chiral}, and thus also of HE to LGE in the classes with CS.  Some further details are discussed in Appendix \ref{app:Classes with CS}.

In the symmetry classes $\mathbb{S}$ with TRS, we prove the equivalence of LGE to HE by a similar approach.  Indeed, we first show that $U_0$ and $U_1$ are LGE if and only if there is some $U_\text{LG}^{(X)}$ that is locally generated in class $X$ for which
\beq
    U_1 = \mct U_\text{LG}^{(X)\dagger}\mct^{-1}U_0 U_\text{LG}^{(X)},\label{eq:LG_canonical_relation_TRS}
\eeq
and then we show that $U_0$ and $U_1$ are HE if and only if the same condition \eqref{eq:LG_canonical_relation_TRS} holds.  These proofs are similar to the CS case that we just presented (see Appendix \ref{app:Classes with TRS} for details).  We thus obtain the equivalence of LGE and HE in all symmetry classes.

We note here that in the four symmetry classes with both CS and TRS, these symmetries are related to PHS by Eq.~\eqref{eq:CS_as_CT}, and therefore the same unitary $U_\text{LG}^{(X)}$ may appear in both Eqs.~\eqref{eq:LG_canonical_relation_chiral} and~\eqref{eq:LG_canonical_relation_TRS} [c.f. Eq.~\eqref{eq:PHS_unitary}]. 

\section{Classification of local unitaries}\label{sec:Classification of local unitaries}

We now begin a formal topological classification of local unitaries for all symmetry classes and dimensions. Our approach will be to obtain a one-to-one mapping between local unitaries and local, flattened Hamiltonians, at which point the machinery used to classify static Hamiltonians can be used. 

The essential point for this classification is that the one-to-one mapping between unitaries and Hamiltonians preserves stable homotopy: if two unitaries $U_1,U_2$ map to Hamiltonians $H_1,H_2$, then $U_1\sim U_2$ if and only if $H_1\sim H_2$.  In proving this statement, we will use the HE definition of equivalence for unitaries rather than the LGE definition.  Since we showed in Sec. \ref{sec:Proof that LGE and HE are equivalent} that HE and LGE are equivalent to each other, the choice of HE for the proof is merely one of convenience.  The classification that we obtain for unitaries may thus be understood as being based on LGE, i.e., it is a classification by local generatability.

The calculation proceeds as follows.  For each AZ symmetry class of unitaries, we construct a one-to-one mapping to a particular AZ symmetry class of Hamiltonians.  These mappings take one form in the case that the unitaries have CS (classes AIII, BDI, DIII, CII, and CI) and another form in the case that the unitaries do not (classes A, AI, D, AII, and C).  After presenting these mappings, we verify that they preserve homotopy equivalence and stable homotopy equivalence, which then allows us to transfer the known results for the classification of Hamiltonians to obtain the classification of unitaries.

\subsection{Unitaries without chiral symmetry}\label{sec:Unitaries without chiral symmetry}
We begin by defining a mapping from unitaries without CS to flattened Hamiltonians with CS.  Given a unitary $U$ without CS, we define the mapping $U\mapsto H_U$ by Eq.~\eqref{eq:H_U_def}.  As noted there, $H_U$ automatically is a local, flattened Hamiltonian with a CS operator $\mcs'=  \sigma^z \otimes \mathbb{I}$, where $\sigma^z$ acts in the same basis that Eq.~\eqref{eq:H_U_def} is written in.

Next, we construct the inverse mapping to $H_U$.  To do this, we consider the five symmetry classes of Hamiltonians with CS (classes A, BDI, CII, CI, and DIII).  The key point is that any symmetry operators that are present can be fixed to certain canonical forms.  It is then straightforward to read off the inverse mapping and the one-to-one correspondence between symmetry classes of unitaries and symmetry classes of Hamiltonians.

Let us write $\Theta$ for an antiunitary operator that depends on the symmetry class and squares to $\pm\id$.  Without loss of generality, we take the symmetry operators of the four classes of Hamiltonians with PHS and TRS to have the following canonical forms:
\bseq
\begin{align}
    \text{BDI and CII: } \mct'&= \id_2\otimes\Theta \text{ and } \mcc'= \sigma_z\otimes\Theta,\label{eq:BDI and CII forms for Hamiltonians}\\
    \text{CI and DIII: }  \mct'&= \sigma_x\otimes\Theta \text{ and } \mcc'= \sigma_y\otimes\Theta.\label{eq:DIII and CI forms for Hamiltonians}
\end{align}
\eseq
The product $\mcs' = e^{i\delta}\mcc'\mct'= \sigma_z\otimes\id$ is automatically a CS.  (Following the phase convention that we discussed in Sec. \ref{sec:HermitianSymmetries}, we choose $\delta=0$ or $\pi/2$ depending on the symmetry class so that no phase appears in $\mcs'=\sigma_z \otimes\id$.)  While we do not need the precise form of $\Theta$, it may be read off from the ``Hamiltonians'' column of Table \ref{tab:Table of canonical forms} in Appendix \ref{app:Canonical forms for symmetry operators}.

It is readily shown that any flattened Hamiltonian with CS ($\mcs'=\sigma_z\otimes\id$) must take the form of the right-hand side of Eq.~\eqref{eq:H_U_def} for some unique unitary $U$ (see Appendix \ref{app:Further details on the mapping from non-chirally-symmetric unitaries to Hamiltonians}); this defines a mapping from Hamiltonians to unitaries.  The locality of $U$ follows immediately from the locality of the given Hamiltonian.  Thus, we have found the inverse mapping to Eq.~\eqref{eq:H_U_def}.  Note, in particular, that the operator $\Theta$ in Eqs.~\eqref{eq:BDI and CII forms for Hamiltonians}--\eqref{eq:DIII and CI forms for Hamiltonians} acts in the same Hilbert space as $U$.

We can now determine the correspondence between the symmetries of $U$ and the symmetries of $H_U$.  We write an antiunitary operator (acting in the same Hilbert space as $U$) as $\Theta$, and we write the $2\times 2$ identity operator as $\mathbb{I}_2$.  The following equivalences are easily shown (see Appendix \ref{app:Further details on the mapping from non-chirally-symmetric unitaries to Hamiltonians}):
\begin{multline}
    \Theta\text{ is a PHS for }U \iff \mathbb{I}_2\otimes \Theta\text{ is a TRS for }H_U\\
    \iff \sigma_z\otimes\Theta \text{ is a PHS for }H_U, \label{eq:ThetaPHS_equivalences}
\end{multline}
and
\begin{multline}
    \Theta\text{ is a TRS for }U \iff \sigma_x\otimes\Theta \text{ is a TRS for }H_U\\
    \iff \sigma_y \otimes \Theta\text{ is a PHS for }H_U. \label{eq:ThetaTRS_equivalences}
\end{multline}

How do the signs (of the squares) of TRS and PHS for $U$ correspond to the signs for $H_U$?  The correspondence follows immediately from
\begin{multline}
    \mathbb{I}_2 \otimes (\Theta^2) =(\mathbb{I}_2\otimes\Theta)^2 = (\sigma_z\otimes\Theta)^2 = (\sigma_x\otimes\Theta)^2 \\
    = - (\sigma_y\otimes\Theta)^2,\label{eq:sign_correspondence}
\end{multline}
where the final minus sign appears because $\sigma_y$ is pure imaginary.

We thus obtain a one-to-one mapping (under $U\leftrightarrow H_U$) between symmetry classes of unitaries and symmetry classes of Hamiltonians.  By \eqref{eq:ThetaPHS_equivalences} and \eqref{eq:sign_correspondence}, the two classes of unitaries with only PHS ($\mcc$) map to Hamiltonian classes with both TRS ($\mct'= \mathbb{I}_2\otimes\mcc$) and PHS ($\mcc' =\sigma_x\otimes\mcc$), with sign factors determined by $\text{sgn}[\mcc^2] =\text{sgn}[(\mct')^2] =\text{sgn}[(\mcc')^2]$.  Similarly, by \eqref{eq:ThetaTRS_equivalences} and \eqref{eq:sign_correspondence}, the two classes of unitaries with only TRS ($\mct$) also map to Hamiltonian classes with both TRS ($\mct'= \sigma_x\otimes\mct$) and PHS ($\mcc'= \sigma_y\otimes\mct$), but with different sign factors: $\text{sgn}[\mct^2] =\text{sgn}[(\mct')^2] =-\text{sgn}[(\mcc')^2]$.

We have thus obtained the following correspondences between AZ classes of unitaries without CS and AZ classes of Hamiltonians with CS:
\beq
{\rm A}\leftrightarrow{\rm AIII}~~~~~~{\rm AI}\leftrightarrow{\rm CI}~~~~~~{\rm D}\leftrightarrow{\rm BDI}\nonumber\\
{\rm AII}\leftrightarrow{\rm DIII}~~~~~~{\rm C}\leftrightarrow{\rm CII},\label{eq:AZ_shift_no_chiral}
\eeq
We have included here the correspondence A$\leftrightarrow$AIII obtained in Sec. \ref{sec:Classification in class A}.  These correspondences are shown in Table~\ref{tab:symmetry_transformations}.  In passing, we note that going from unitaries to Hamiltonians is equivalent to cyclically moving one row upwards in Table~\ref{tab:periodic} (considering the complex and real classes independently).

\begin{table*}
\be
\renewcommand\arraystretch{1.4}
\begin{array}{c|ccc||c|c|ccc|c}
\hline\hline
\multicolumn{4}{c||}{U} & \multicolumn{6}{c}{H_U}\\
\hline\hline
\mbox{AZ class} & \mct & \mcc & \mcs & H_U & {\rm Symmetry~operators} & \mct' & \mcc' & \mcs' & \mbox{AZ class}\\
\hline\hline
{\rm A} & 0 & 0 & 0 & U\hat{\oplus}U^\dagger & \mcs'=\sigma_z \otimes\id & 0 & 0 & 1 & {\rm AIII}\\
{\rm AIII} & 0 & 0 & 1 & \mcs U & {\rm None} & 0 & 0 & 0 & {\rm A}\\
\hline
{\rm AI} & + & 0 & 0 & U\hat{\oplus}U^\dagger & \mct'=\sigma_x\otimes\mct,~\mcc'=\sigma_y\otimes\mct,~\mcs'=\sigma_z\otimes\id & + & - & 1 & {\rm CI}\\
{\rm BDI} & + & + & 1 & \mcs U & \mct'=\mcc & + & 0 & 0 & {\rm AI}\\
{\rm D} & 0 & + & 0 & U\hat{\oplus}U^\dagger & \mct'=\id_2\otimes\mcc,~\mcc'=\sigma_z\otimes\mcc,~\mcs'=\sigma_z \otimes\id& + & + & 1 & {\rm BDI}\\
{\rm DIII} & - & + & 1 & \mcs U & \mcc'=\mcc & 0 & + & 0 & {\rm D}\\
{\rm AII} & - & 0 & 0 & U\hat{\oplus}U^\dagger & \mct'=\sigma_x\otimes\mct,~\mcc'=\sigma_y\otimes\mct,~\mcs'=\sigma_z\otimes\id & - & + & 1 & {\rm DIII}\\
{\rm CII} & - & - & 1 & \mcs U & \mct'=\mcc & - & 0 & 0 & {\rm AII}\\
{\rm C} & 0 & - & 0 & U\hat{\oplus}U^\dagger & \mct'=\id_2\otimes\mcc,~\mcc'=\sigma_z\otimes\mcc,~\mcs'=\sigma_z \otimes\id & - & - & 1 & {\rm CII}\\
{\rm CI} & + & - & 1 & \mcs U & \mcc'=\mcc & 0 & - & 0 & {\rm C}
\end{array}
\ee
\caption{Summary of the mapping between a local unitary $U$ and a local, flattened Hamiltonian $H_U$ for each symmetry class. We use the shorthand $U\hat{\oplus}U^\dagger$ to represent the doubled Hamiltonian obtained from $U$ and $U^\dagger$ through Eq.~\eqref{eq:H_U_def}.  See main text for details. \label{tab:symmetry_transformations}}
\end{table*}

\subsection{Unitaries with chiral symmetry}\label{sec:Unitaries with chiral symmetry}
We repeat the steps of the previous section, now using a different mapping from unitaries to flattened Hamiltonians.  We construct a one-to-one mapping between symmetry classes of unitaries with CS and symmetry classes of Hamiltonians without CS.

When CS is present, we may bring it to the form \eqref{eq:CS_canonical_form} without loss of generality.  Given a unitary $U$ in a symmetry class with CS, we then define
\bequ
    H_U = \mcs U.\label{eq:HU_def_chiral}
\eequ

The locality of $H_U$ follows from the locality of $U$.  Furthermore, the CS property of $U$, the unitarity of $\mcs$ and $U$, and Eq.~\eqref{eq:CS_inverse} imply that $H_U$ is indeed Hermitian and flattened.

However, $\mcs$ is not a CS for $H_U$, since
\beq
\mcs H_U\mcs ^{-1}=U\mcs=\mcs U^\dagger,\label{eq:mcs not a CS}
\eeq
which is not generally equal to $-\mcs U$.  (We ignore the ``fine-tuned'' set of unitaries that satisfy $U=-U^\dagger$, with no effect on our classification.)  Thus, the mapping \eqref{eq:HU_def_chiral} sends unitaries with CS to Hamiltonians without CS.  The class of unitaries with only CS (class AIII) evidently maps to the class of Hamiltonians with no symmetry (class A).

Next, we construct the inverse mapping to $H_U$ and find the one-to-one correspondence between symmetry classes of unitaries with CS and symmetry classes of Hamiltonians without CS.  As in the previous section, the key point is that any symmetry operators that are present can be fixed to certain canonical forms.  We also exploit the fact that in any given symmetry class $\mathbb{S}$, we are free to choose one set of canonical forms for the symmetry operators of \emph{unitaries} in class $\mathbb{S}$ and another set of canonical forms for \emph{Hamiltonians} in class $\mathbb{S}$.

Before constructing the inverse mapping, we first explain a technical point regarding the number of generalized orbitals per lattice site (call this number $\alpha$).  Since $U$ has CS, $\alpha$ is always even; furthermore, in class CII, $\alpha$ is a multiple of four (c.f. the canonical form for PHS in CII below).  The same value of $\alpha$ is then obtained for $H_U$ by construction.  However, some Hamiltonian symmetry classes without CS (in particular, classes A and AI) can have $\alpha$ odd or even, while the remaining classes have $\alpha$ even, but not necessarily a multiple of four.  Thus, the inverse mapping can generally only be defined on a \emph{subset} of Hamiltonians.  In three of the Hamiltonian classes without CS, we therefore restrict to the subspace of Hamiltonians with $\alpha$ satisfying
\bequ
    \alpha \text{ is a multiple of }
    \begin{cases}
        2 & \text{classes A and AI},\\
        4 & \text{class AII}.
    \end{cases}\label{eq:alpha conditions}
\eequ
In the remaining two classes (C and D), no restriction is necessary, as we see below.

In Appendix \ref{app:Classification with restricted number of orbitals}, we show that the topological classification of Hamiltonians is the same whether or not the restrictions \eqref{eq:alpha conditions} are imposed.  Thus, \eqref{eq:alpha conditions} does not affect the mapping that we establish between symmetry classes of unitaries and of Hamiltonians.

We proceed to construct the inverse mapping.  The idea is to \emph{define} an operator $\mcs$ by Eq.~\eqref{eq:CS_canonical_form}; then, although $\mcs$ is not a CS for $H$, we map a given $H$ from a class without CS to $U_H= \mcs H$, which is a local, chirally symmetric unitary with CS operator $\mcs$.  (That $\mcs$ is a CS for $U_H$ follows from $H$ being Hermitian, and the unitarity of $U_H$ follows from $H$ being Hermitian and flattened.)  Thus, the inverse mapping to Eq.~\eqref{eq:HU_def_chiral} is
\beq
    U_H = \mcs H.\label{eq:HU_inverse_chiral}
\eeq
The remaining task of this section is to verify that we can indeed define $\mcs$ by Eq.~\eqref{eq:CS_canonical_form} and to determine the correspondence of symmetries under the mapping $U\leftrightarrow H_U$.

For class A Hamiltonians with an even number of orbitals [c.f. \eqref{eq:alpha conditions}], we make an arbitrary choice of half of the orbitals and declare that $\mcs$ acts as $1$ on these and as $-1$ on the remainder, i.e., we define $\mcs$ by Eq.~\eqref{eq:CS_canonical_form}.  For the remaining four classes of Hamiltonians without CS (classes AI, AII, D, and C), we fix the symmetry operators to the following canonical forms without loss of generality (as in, e.g., Table I of Ref.~\cite{RoyPeriodic2017}):
\bequ
    \text{AI and AII: } \mct' = \id_2\otimes\Theta,\label{eq:AI and AII form for Hamiltonians}
\eequ
where $\Theta$ is some antiunitary operator that depends on the symmetry class [and where the identity matrix is $2\times 2$ due to \eqref{eq:alpha conditions}], and 
\bequ
    \text{D and C: } \mcc'=\Theta\otimes\id,\label{eq:D and C form for Hamiltonians}
\eequ
where $\Theta$ is some $2\times 2$ antiunitary operator that depends on the symmetry class and anticommutes with $\sigma_z$.  We then define $\mcs$ by Eq.~\eqref{eq:CS_canonical_form}.  While the explicit form of $\Theta$ is not needed, we note that it is $\Theta= \mathcal{K}$ in AI, $\mathcal{K} \sigma_x$ in D, and $\mathcal{K}\sigma_y$ in AII and C \cite{RoyPeriodic2017}.

We have thus shown that Eq.~\eqref{eq:HU_def_chiral} is a one-to-one mapping with inverse given by $U = \mcs H_U$, where $\mcs$ is given by \eqref{eq:CS_canonical_form}.  Since we have already seen that class AIII unitaries map to class A Hamiltonians, it follows that $U\leftrightarrow H_U$ is a one-to-one mapping between these classes (provided that the Hamiltonians are restricted to those with even $\alpha$).

To determine the correspondence, under $U\leftrightarrow H_U$, of the remaining symmetry classes of unitaries with CS to symmetry classes of Hamiltonians without CS, we first fix the PHS operator of $U$ to a certain canonical form in each symmetry class.  (The form of the TRS operator will not be needed.)  Without loss of generality, we take the PHS operator $\mcc$ to be
\bequ
    \text{BDI and CII: } \mcc= \id_2\otimes\Theta,\label{eq:BDI and CII form for unitaries}
\eequ
where $\Theta$ is some antiunitary operator that depends on the symmetry class, and 
\bequ
    \text{DIII and CI: }  \mcc= \Theta\otimes\id,\label{eq:DIII and CI form for unitaries}
\eequ
where $\Theta$ is some $2\times 2$ antiunitary operator that depends on the symmetry class and anticommutes with $\sigma_z$.  While we do not need the precise form of $\Theta$, nor the form of the TRS operator, they may be read off from the ``Unitaries'' column of Table \ref{tab:Table of canonical forms} in Appendix \ref{app:Canonical forms for symmetry operators}.

We emphasize that there is no need for Eqs.~\eqref{eq:BDI and CII form for unitaries} and \eqref{eq:DIII and CI form for unitaries} to match the forms for PHS that we used earlier for Hamiltonians in these same symmetry classes [Eqs.~\eqref{eq:BDI and CII forms for Hamiltonians} and \eqref{eq:DIII and CI forms for Hamiltonians}].  Here we are fixing the canonical forms in symmetry classes of \emph{unitaries}.

Finally, we make use of the close similarity between Eqs.~\eqref{eq:AI and AII form for Hamiltonians} and \eqref{eq:D and C form for Hamiltonians} and Eqs.~\eqref{eq:BDI and CII form for unitaries} and \eqref{eq:DIII and CI form for unitaries}.  In all four cases, the symmetry operator ($\mct',\mcc'$, or $\mcc$) is an antiunitary operator $\Lambda$ that either commutes or anticommutes with $\mcs$ [c.f. Eq.~\eqref{eq:CS_canonical_form}].  (Achieving this was the point of fixing the canonical forms in the way that we did.)  We cover both possibilities by showing the following two equivalences: if $[\Lambda,\mcs]=0$, then
\bequ
    \Lambda\text{ is a PHS for }U \iff \Lambda\text{ is a TRS for }H_U, \label{eq:ThetaTRS_equivalence_chiral}
\eequ
and if instead $\{\Lambda,\mcs\}=0$, then
\bequ
    \Lambda\text{ is a PHS for }U \iff \Lambda \text{ is a PHS for }H_U.\label{eq:ThetaPHS_equivalence_chiral}
\eequ
These equivalences follow immediately from
\bequ
\Lambda H_U\Lambda^{-1} = 
\begin{cases}
    S \Lambda U \Lambda^{-1} & [\Lambda,\mcs]=0,\\ 
    -S \Lambda U \Lambda^{-1} & \{\Lambda,\mcs\}=0.
\end{cases}
\eequ

Comparing Eqs.~\eqref{eq:BDI and CII form for unitaries} and \eqref{eq:AI and AII form for Hamiltonians}, we see that in classes BDI and CII of unitaries (or, equivalently, classes AI and AII for Hamiltonians), Eq.~\eqref{eq:ThetaTRS_equivalence_chiral} applies, and hence $\mcc=\mct'$.  The sign factors are then related by $\text{sgn}[\mcc^2]=\text{sgn}[(\mct')^2]$.  Similarly, comparing Eqs.~\eqref{eq:DIII and CI form for unitaries} and \eqref{eq:D and C form for Hamiltonians}, we see that in classes DIII and CI of unitaries (or, equivalently, classes D and C of Hamiltonians), Eq.~\eqref{eq:ThetaPHS_equivalence_chiral} applies, and hence $\mcc=\mcc'$ and $\text{sgn}[\mcc^2]=\text{sgn}[(\mcc')^2]$. 

Thus, the one-to-one mapping $U\leftrightarrow H_U$ connects symmetry classes in the following way: unitary symmetry classes with PHS and TRS map to Hamiltonian classes with either TRS only or PHS only, with the sign factor of the Hamiltonian symmetry being the same as the sign factor of the PHS for the unitary.  In other words, the CS and TRS of $U$ are lost, and the PHS becomes either TRS or PHS for $H_U$.  The mapping between class AIII unitaries and class A Hamiltonians also fits this pattern (the CS of $U$ is lost).

Thus, we have obtained the following correspondence between classes of unitaries with CS and classes of Hamiltonians without CS:
\beq
{\rm AIII}\leftrightarrow{\rm A}~~~~~~{\rm BDI}\leftrightarrow{\rm AI}~~~~~~{\rm CII}\leftrightarrow{\rm AII}\nonumber\\
{\rm DIII}\leftrightarrow{\rm D}~~~~~~{\rm CI}\leftrightarrow{\rm C}.\label{eq:AZ_shift_chiral}
\eeq
These correspondences are shown in Table~\ref{tab:symmetry_transformations}.  The same cyclic pattern noted below~\eqref{eq:AZ_shift_no_chiral} also holds here.

\subsection{Topological classification of local unitaries\label{sec:top_class_unitaries}}
We can now present the classification of local unitaries.  The classification will be according to the stable equivalence relation $\sim$, which we previously defined in class A [Eq.~\eqref{eq:def_stable_equivalence_of_unitaries}]; we use the same definition in all other symmetry classes with the additional requirement that the two trivial unitaries $U_0^{(0)}$ and $U_1^{(0)}$ belong to the given symmetry class.

The key point is that the one-to-one mappings that we have used between unitaries and Hamiltonians preserve equivalence up to stable homotopy.  In particular, within any symmetry class $\mathbb{S}$ of unitaries, the equivalence \eqref{eq:mapping preserves stable homotopy} (that we discussed earlier in the class A case) holds.  The three symmetry classes referred to in \eqref{eq:alpha conditions} require special treatment; in these cases, the relation $\sim$ between the Hamiltonians $H_{U_0}$ and $H_{U_1}$ is understood to be replaced by a relation $\overset{r}{\sim}$ which restricts the number of generalized orbitals according to \eqref{eq:alpha conditions} (see Appendix \ref{app:Classification with restricted number of orbitals}).  The equivalence \eqref{eq:mapping preserves stable homotopy} may be verified essentially by inspection of the mappings \eqref{eq:H_U_def} and \eqref{eq:HU_def_chiral}; we present the explicit verification in Appendix \ref{app:Mapping between unitaries and Hamiltonians preserves stable homotopy}.

In this way, the topological classification of local unitaries  \emph{directly follows} from the existing topological classification of gapped Hamiltonians.  From the correspondences \eqref{eq:AZ_shift_no_chiral} and \eqref{eq:AZ_shift_chiral}, we read off Table~\ref{tab:periodic_unitary}, the periodic table for local unitaries.

\begin{table}
\be
\begin{array}{c|ccc|cccccccc}
\hline
\hline
\mbox{AZ class} & \mct & \mcc & \mcs & d=0 & 1 & 2 & 3 & 4 & 5 & 6 & 7\\
\hline\hline
\mbox{A} & 0 & 0 & 0 & 0 & \zbb & 0 & \zbb & 0 & \zbb & 0 & \zbb\\
\mbox{AIII} & 0 & 0 & 1 & \zbb & 0 & \zbb & 0 & \zbb & 0 & \zbb & 0\\
\hline
\mbox{AI} & + & 0 & 0 & 0 & 0 & 0 & \zbb & 0 & \zbb_2 & \zbb_2 & \zbb \\
\mbox{BDI} & + & + & 1 & \zbb & 0 & 0 & 0 & \zbb & 0 & \zbb_2 & \zbb_2 \\
\mbox{D} & 0 & + & 0 & \zbb_2 & \zbb & 0 & 0 & 0 & \zbb & 0 & \zbb_2 \\
\mbox{DIII} & - & + & 1 & \zbb_2 & \zbb_2 & \zbb & 0 & 0 & 0 & \zbb & 0 \\
\mbox{AII} & - & 0 & 0 & 0 & \zbb_2 & \zbb_2 & \zbb & 0 & 0 & 0 & \zbb \\
\mbox{CII} & - & - & 1 & \zbb & 0 & \zbb_2 & \zbb_2 & \zbb & 0 & 0 & 0 \\
\mbox{C} & 0 & - & 0 & 0 & \zbb & 0 & \zbb_2 & \zbb_2 & \zbb & 0 & 0 \\
\mbox{CI} & + & - & 1 & 0 & 0 & \zbb & 0 & \zbb_2 & \zbb_2 & \zbb & 0 \\
\hline\hline
\end{array}
\ee
\caption{Periodic table for local unitaries. The leftmost four columns define the AZ symmetry classes, as described in the main text and in Table~\ref{tab:periodic}. The rightmost eight columns indicate the topological classification for local unitaries from each symmetry class in dimension $d$. Note that this table is a permutation of the periodic table for topological Hamiltonians.\label{tab:periodic_unitary}}
\end{table}

We now comment in more detail on the relation of our work to the work of Higashikawa \emph{et al.} \cite{HigashikawaFloquet2019} mentioned in the Introduction.  Reference \cite{HigashikawaFloquet2019} classifies translation-invariant bulk unitaries that can be obtained from block-diagonal Floquet operators by projection onto one of the blocks.  The nontrivial unitaries obtained in this way are gapless. The classification table obtained in Ref.~\cite{HigashikawaFloquet2019} is thus of possible bulk gapless states and is obtained by mapping unitaries to Hamiltonians using maps similar to the ones used here. Their classification relies on the plausible conjecture that every nontrivial unitary obtained by mapping nontrivial Hamiltonians can be realized by a band projection of a (trivial) Floquet unitary. We note also that our paper, in contrast to Ref.~\cite{HigashikawaFloquet2019}, includes equivalence up to local generatability and does not assume translation invariance.

\section{Examples of Nontrivial Local Unitaries\label{sec:examples}}
In this section, we provide several examples of topologically nontrivial unitaries.  These examples illustrate how the one-to-one mappings that we introduced between symmetry classes (of unitaries and Hamiltonians) can be used to diagnose the topology of Hamiltonians and also to find explicit formulas for topological invariants of unitaries.

We provide example unitaries in all topologically nontrivial symmetry classes in one dimension and in one of the nontrivial symmetry classes in two dimensions.  For simplicity, we consider only the translation-invariant case.

\subsection{One dimension}

From Table \ref{tab:periodic_unitary}, we recall that the nontrivial symmetry classes are A, D, AII, DIII, and C.  Of these, only DIII has CS.  Below, we first consider the four classes without CS, presenting the class A case in some detail and then summarizing the results for the others; then, we consider class DIII. 

\subsubsection{Unitaries without chiral symmetry}
We define two class A unitaries in momentum space by
\bseq
\begin{align}
U_\text{triv}^{\rm(A)}(k)& =
\bpmat
    0 & 1 \\
    1 & 0
\epmat,\label{eq:Utriv_A}\\
U_\text{top}^{\rm(A)}(k)&=e^{ik}\label{eq:Utop_A}.
\end{align}
\eseq
In the first case, each lattice site has two orbitals, and the unitary enacts a hop between the two orbitals on each site.  In real space, the unitary consists of repeating blocks of the Pauli operator $\sigma_x$, i.e., we have
\beq
U_\text{triv}^{\rm(A)} = \bigoplus_i\left(\begin{array}{cc}
0 & 1 \\
1 & 0
\end{array}\right)_i,\label{eq:Utriv_A_real_space}
\eeq
which is clearly a local operator because it acts separately on each site. It is also locally generated; this may be verified by noting
\beq
U_\text{triv}^{\rm(A)} = \exp\left[-i\frac{\pi}{2}\bigoplus_i\left(\begin{array}{cc}
-1 & 1 \\
1 & -1
\end{array}\right)_i\right],
\eeq
in which the Hamiltonian in the exponent also acts only on site (and is thus local).

The second unitary, Eq.~\eqref{eq:Utop_A}, acts on a lattice with one orbital per site as a translation by one unit to the left, with real-space matrix elements
\beq
U_{\text{triv},ij}^{\rm(A)}= \delta_{i,j+1}.\label{eq:Utop_A_real_space}
\eeq
Since these matrix elements are strictly zero unless $i=j+1$, this unitary is also local. However, it is not locally generated, as may be verified naively by studying the matrix elements of its matrix logarithm as a function of system size. More fundamentally, this property stems from the fact that translation is anomalous on a one-dimensional lattice \cite{KitaevAnyons2006}.

\begin{figure}
\includegraphics[scale=0.17]{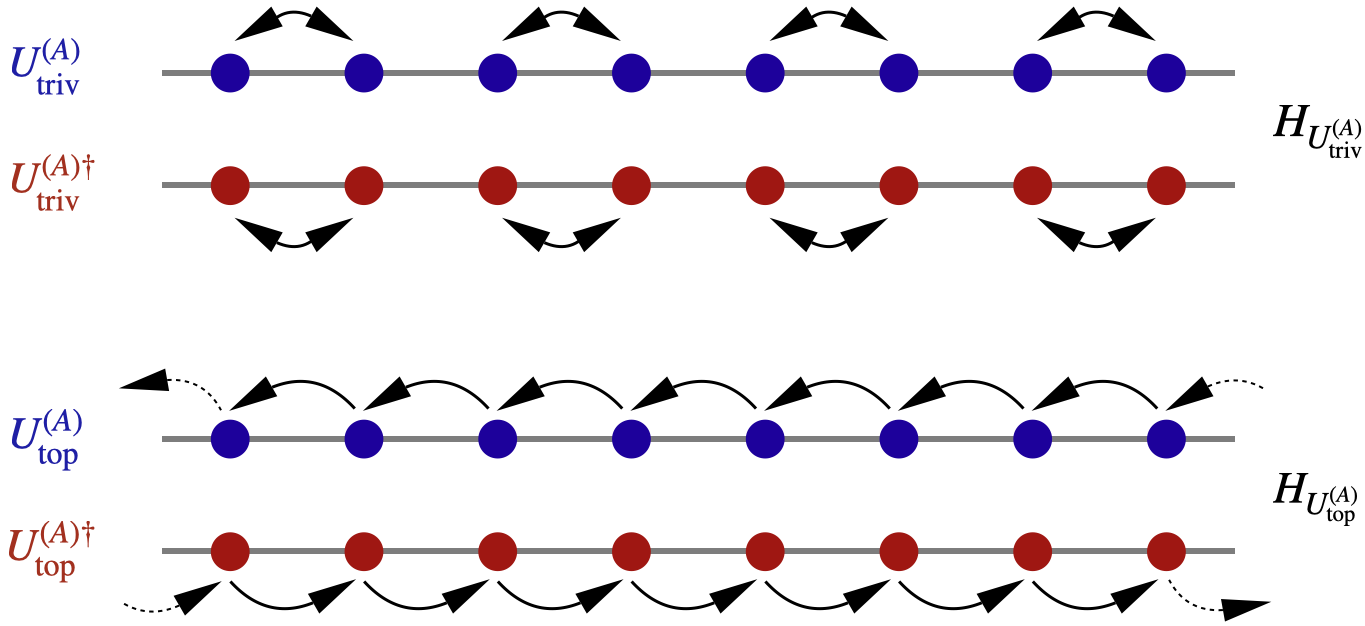}
\caption{Illustration of the actions of the two example class A unitaries \eqref{eq:Utriv_A} and \eqref{eq:Utop_A} and of the class AIII Hamiltonians \eqref{eq:HUtriv_A} and \eqref{eq:HUtop_A} that they map to under \eqref{eq:H_U_def}.\label{fig:HU_classA_1d}}
\end{figure}

Following the approach of Sec.~\ref{sec:Unitaries without chiral symmetry}, we use Eq.~\eqref{eq:H_U_def} to map these unitaries to class AIII Hamiltonians, yielding
\bseq
\begin{align}
H_{U_\text{triv}^{\rm(A)}}(k)&=\left(\begin{array}{cccc}
0 & 0 & 0 & 1 \\
0 & 0 & 1 & 0 \\
0 & 1 & 0 & 0 \\
1 & 0 & 0 & 0
\end{array}\right),\label{eq:HUtriv_A}\\
H_{U_\text{top}^{\rm(A)}}(k)&=\left(\begin{array}{cc}
0 & e^{ik} \\
e^{-ik} & 0 
\end{array}\right).\label{eq:HUtop_A}
\end{align}
\eseq
Physically, the mapping to Hamiltonians may be interpreted as adding a sublattice degree of freedom to the system, where $H_{U_{\text{triv}}^{(\rm A)}}$ then performs intracell hops, while $H_{U_{\text{top}}^{(\rm A)}}$ performs hopping in opposite directions for each sublattice (see Fig.~\ref{fig:HU_classA_1d}). 

By calculating the class~AIII topological invariant for these Hamiltonians \cite{Mondragon-ShemTopological2014}, it may be verified that $U_\text{triv}^{\rm(A)}$ is trivial, while $U_\text{top}^{\rm(A)}$ is topological with an invariant of $1$. Indeed, the invariant for these Hamiltonians is simply the winding number of the off-diagonal block, i.e., of the unitary in Eq.~\eqref{eq:H_U_def}. We have thus connected the known invariant in the Hamiltonian case to the known invariant in the unitary case (recall that for a translation-invariant, 1D local unitary in class A, the invariant is the winding number~\cite{KitaevAnyons2006,GrossIndex2012}). While we have only presented two simple examples, the connection is more general.

A similar picture also holds for  nontrivial unitaries in classes D and C, which can be mapped to Hamiltonians with CS and a winding number invariant. In class~AII, a local unitary is mapped to a Hamiltonian in class DIII [c.f. \eqref{eq:AZ_shift_no_chiral}].  The corresponding $\zbb_2$ topological invariant in this case (for both unitaries and Hamiltonians) may loosely be thought of as the winding number taken modulo $2$. More formally, a constrained version of the Fu-Kane invariant may be used instead \cite{TeoTopological2010}.

\subsubsection{Unitaries with chiral symmetry}
The only remaining nontrivial case in 1D is class DIII, which maps to class D for Hamiltonians [c.f. \eqref{eq:AZ_shift_chiral}].  Topological Hamiltonians in class~D are exemplified by the well-known $p$-wave superconducting chain \cite{KitaevUnpaired2001}.  We take this as our starting point and use our one-to-one mapping to obtain a nontrivial unitary in class~DIII.

The Hamiltonian for the $p$-wave superconductor may be written as~\cite{KitaevUnpaired2001}
\beq
\hat{H}_{\rm K}&=&-\mu\sum_j\left[ c^\dagger_{j}c^\ph_j-\frac{1}{2}\right]\nonumber\\
&&-\frac{1}{2}\sum_j\left[tc^\dagger_{j}c^\ph_{j+1}+\Delta e^{i\phi}c^\ph_{j}c^\ph_{j+1}+\text{H.c.}\right],
\eeq
where $\mu$ is the chemical potential, $t$ is the hopping parameter, and $\Delta e^{i\phi}$ is the superconducting pairing strength. We consider the two easily solvable (and flattened) points of this model as examples of trivial and topological flattened Hamiltonians from class D. In BdG form and in momentum space, these fixed points have the Hamiltonians
\bseq
\begin{align}
H^{\rm(D)}_{\rm triv}(k)&= \left(\begin{array}{cc}
1 & 0 \\
0 & -1
\end{array}\right),\label{eq:HD_triv}\\
H^{\rm(D)}_{\rm top}(k)&=\left(\begin{array}{cc}
-\cos(k) & i\sin(k) \\
-i\sin(k) & \cos(k)
\end{array}\right),\label{eq:HD_top}
\end{align}
\eseq
which each (it may be verified) square to the identity matrix and satisfy PHS with $\mcc=\mathcal{K}\sigma_x$.  (These fixed point Hamiltonians also have additional symmetries, but they do not affect the arguments below.)  It may also be verified (e.g., by transforming to the Majorana fermion basis and calculating the Pfaffian \cite{KitaevUnpaired2001}) that these Hamiltonians have invariants $0$ and $1$, respectively.

We now map these Hamiltonians to unitaries using Eq.~\eqref{eq:HU_inverse_chiral}.    Following the approach described there [c.f. also Eq.~\eqref{eq:D and C form for Hamiltonians}], we define the CS operator as $S=\sigma_z$.  Then these two Hamiltonians map to
\bseq
\begin{align}
U_\text{triv}^{(\rm DIII)}(k)&=\left(\begin{array}{cccc}
1&0 \\
0 & 1
\end{array}\right),\\
U_{\text{top}}^{(\rm DIII)}(k)&=\left(\begin{array}{cccc}
-\cos(k) & i\sin(k) \\
i\sin(k) & -\cos(k)
\end{array}\right).
\end{align}
\eseq
As a check, we note that $U_\text{triv}^{(\rm DIII)}(k)$ is indeed trivial because it is an identity matrix, while the nontriviality of $U_\text{top}^{(\rm DIII)}$ is consistent with its gapless quasienergy spectrum (which is readily calculated).

\subsection{Two examples in two dimensions}
We now present two examples of topologically nontrivial unitaries in two-dimensional class AIII.  We start from nontrivial Hamiltonians $H^{(\rm{A})}(\bk)$ from class A and use the inverse mapping from Hamiltonians to unitaries to obtain topologically nontrivial $U^{(\rm AIII)}(\bk)$.   For $H^{(\rm{A})}(\bk)$, we consider two different models of two-band Chern insulators (from Refs.~\cite{BernevigTopological2013,RoyInteger2006}), specializing the values of the parameters such that the Hamiltonian is topologically nontrivial.  In both cases, $H^{(\rm{A})}(\bk) =\bm{\sigma}\cdot \mathbf{n}(\bk)$ for some three-vector $\mathbf{n}(\bk)$.

Following the approach of Sec.~\ref{sec:Unitaries with chiral symmetry}, we define a CS operator by $\mcs = \sigma_z$ (acting on the two bands). Then, applying the mapping \eqref{eq:HU_inverse_chiral} to $H^{(\rm A)}(k)$, we obtain the following unitary:
\bequ
    U^{(\rm AIII)}(\bk) = \frac{1}{|\mathbf{n}(\bk)|}\left(n_z \id -i n_y \sigma_x + i n_x \sigma_y \right),\label{eq:2d unitary}
\eequ
where $|\mathbf{n}(\bk)|$ appears due to the flattening of $H^{(\rm{A})}(\bk)$.

\begin{figure}
\vspace{1mm}
\includegraphics[scale=0.17]{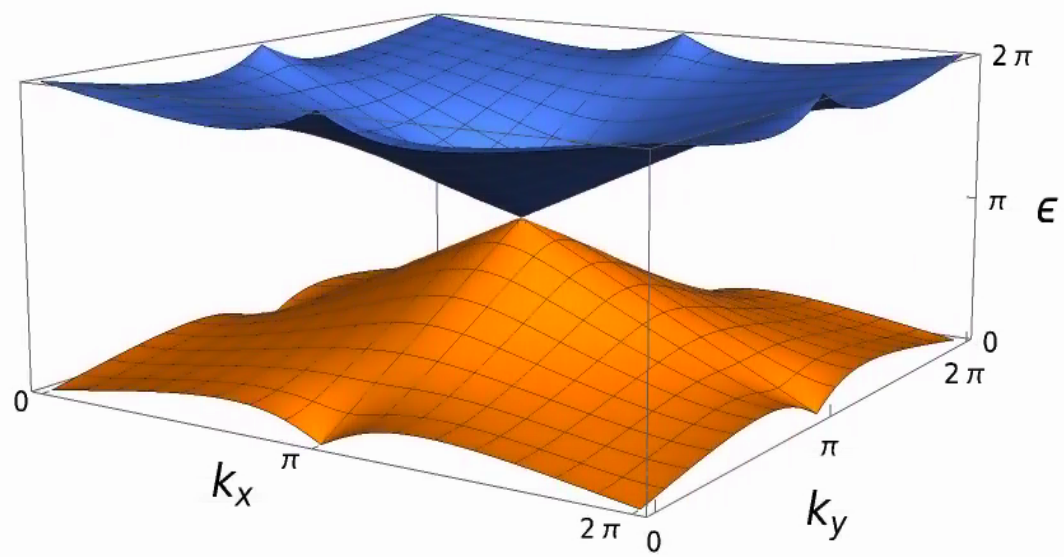}
\caption{ Three-dimensional view of the quasienergy spectrum of the unitary $U^{(\rm{AIII})}(k)$ obtained from the Chern insulator Hamiltonian in Ref.~\cite{BernevigTopological2013}.
 \label{fig:AIII_spectra}}
\end{figure}

For our first example, we let $H^{(\rm{A})}(\bk)$ be the Hamiltonian given by Eq. (8.17) in Ref.~\cite{BernevigTopological2013}.  Setting the parameters there to $M=-3$ and $B=-1$ yields $\mathbf{n}(\bk)= (\sin k_x, \sin k_y, 1+ \cos k_x + \cos k_y)$.  Then, the Chern number of $H^{(\rm{A})}(\bk)$ is $1$ \cite{BernevigTopological2013}.  The numerical computation of the quasienergy spectrum of the unitary given by Eq.~\eqref{eq:2d unitary} reveals a Dirac cone (Fig.~\ref{fig:AIII_spectra}). 

For our second example, we let $H^{(\rm{A})}(\bk)$ be the Hamiltonian given by Eq. (1) of Ref.~\cite{RoyInteger2006}.  Setting the parameters there to $a=t=1,\Phi=\pi/2,\gamma_1=0$, and $\gamma_2=\pi$, we obtain $\mathbf{n}(\bk) = (2(\sqrt{2}\cos k_x + \cos k_y), -\sqrt{2}\cos k_x, M + 2 t_2 \sin k_x \sin k_y)$, where $t_2$ and $M$ are real parameters \footnote{Here we have corrected typos in Eq. (1) of Ref.~\cite{RoyInteger2006}: the second left parentheses in the first line and the last right parentheses in the second line should be deleted.}.  The Chern number of $H^{(\rm{A})}(\bk)$ is then equal to $1$ if $|M| < t_2$ and zero otherwise \cite{RoyInteger2006}.  Thus, Eq.~\eqref{eq:2d unitary} with $|M|<t_2$ provides another example of a nontrivial unitary in two-dimensional class AIII.

\section{Conclusion\label{sec:conclusion}}
In this paper, we presented a topological classification of noninteracting, local unitaries in all dimensions and symmetry classes based on equivalence up to local generatability.  To achieve this, we defined the notion of equivalence up to local generatability in the presence of symmetry and showed that it is equivalent to homotopy equivalence.  We then used a one-to-one, homotopy-preserving mapping between unitaries and Hamiltonians to read off, from the existing classification of Hamiltonians, the classification of unitaries up to HE, and thus also up to LGE.  (More precisely we used a more general notion of stable HE based on K-theoretic stable homotopy from the translation invariant case.)

Our paper opens several natural directions for future study.  The one-to-one correspondence between unitaries and static Hamiltonians can be used both to generate more examples of topologically nontrivial unitaries and to find explicit formulas for topological invariants (by transforming a known invariant from the static case).  We presented some examples of this approach in Sec. \ref{sec:examples}, but there are many more AZ symmetry classes to explore in this way.  It would be interesting to see if the resulting real-space invariants are locally computable, as in the nonsymmetric 1D case studied in Ref.~\cite{GrossIndex2012}.   Another follow-up to our paper would be to determine a precise bulk-boundary correspondence between $(d+1)$-dimensional Floquet operators restricted to the edge and $d$-dimensional local unitaries. 

While we have focused throughout on the noninteracting case, we note that progress has been made recently in related problems in the interacting case, including the construction of nontrivial quantum cellular automata in dimensions higher than one \cite{HaahNontrivial2023} and the classification of translation-invariant Clifford quantum cellular automata \cite{HaahTopological2022} and Clifford Floquet circuits \cite{GeikoHomotopy2023}.  It would be interesting to see if the ideas developed in this paper have any extension to the interacting case.  It is conceivable that enlarging the set of locally generated unitaries to include interacting unitaries could destabilize our classification.

\begin{acknowledgments}
This material is based upon work supported by the National Science Foundation under Grant No. DMR-1455368.  This research was supported by funds from the UC National Laboratory Fees Research Program of the University of California, Grant Number LFR-20-653926.  A.~B.~C acknowledges financial support from the Joseph P. Rudnick Prize Postdoctoral Fellowship (UCLA).
\end{acknowledgments}

\appendix
\begin{widetext}

\section{Properties of the composition operation}
\subsection{Composition preserves symmetry}\label{app:Composition preserves symmetry}
We verify that the composition operation defined in Eq.~\eqref{eq:composition_with_LG} preserves any symmetries present.

First, consider PHS.  From Eq.~\eqref{eq:PHS_ham} and the antiunitarity of $\mcc$, we read off (for any times $t_1,t_2$)
\beq
    \mcc\tord\exp\left(-i\int_{t_1}^{t_2}dt\ H(t) \right)\mcc^{-1} = \tord \exp\left(-i\int_{t_1}^{t_2}dt\ H(t) \right). 
\eeq
Thus,
\beq
    \mcc( U\star U_\text{LG}) \mcc^{-1} =  \left[\tord\exp\left(-i\int_{1/2}^1 dt\ H(t)  \right) \right] \mcc U \mcc^{-1} \tord\left(-i\int_0^{1/2}dt\  H(t) \right) = U\star U_\text{LG},
\eeq
by Eq.~\eqref{eq:PHS_unitary}.  Thus, $U\star U_\text{LG}$ has PHS.

For the cases of CS and TRS, and also for later calculations, it is useful to show the following two identities.  Consider a symmetry class $\mathbb{S}$ with CS, TRS, or both, and define the symmetry class $X$ by Eq.~\eqref{eq:X}.  If $\mathbb{S}$ has CS, then the composition of any $U$ in $\mathbb{S}$ with any $U_\text{LG}$ that is locally generated in $\mathbb{S}$ [by some $H(t)$] may be written as
\beq
    U\star U_\text{LG} = \mcs U_\text{LG}^{(X)\dagger}\mcs^{-1} U U_\text{LG}^{(X)},\label{eq:composition_CS}
\eeq
where $U_\text{LG}^{(X)}$ is locally generated in $X$ by $H^{(X)}(t) = \frac{1}{2}H(\frac{1}{2}t)$.  Similarly, if $\mathbb{S}$ has TRS, then the composition of any $U$ in $\mathbb{S}$ with any $U_\text{LG}$ that is locally generated in $\mathbb{S}$ [by some $H(t)$] may be written as
\beq
    U\star U_\text{LG} = \mct U_\text{LG}^{(X)\dagger}\mct^{-1} U U_\text{LG}^{(X)},\label{eq:composition_TRS}
\eeq
where $U_\text{LG}^{(X)}$ is locally generated in $X$ by $H^{(X)}(t) = \frac{1}{2}H(\frac{1}{2}t)$.

To show these two identities, we start by showing
\bseq
\begin{align}
    \tord\exp\left(-i\int_{1/2}^1dt\ H(t) \right)&= \mcs \left[\tord\exp\left(-i\int_0^{1/2}dt\ H(t) \right) \right]^\dagger\mcs^{-1},\label{eq:identity_CS}\\
    \tord\exp\left(-i\int_{1/2}^1 dt\ H(t) \right)&= \mct \left[\tord\exp\left(-i\int_0^{1/2}dt\ H(t) \right) \right]^\dagger\mct^{-1},\label{eq:identity_TRS}
\end{align}
\eseq
which are obtained as follows.  We relabel $t\to1-t$ and use Eq.~\eqref{eq:CS_Hamiltonian_floquet} to write the left-hand side of \eqref{eq:identity_CS} as  $\antitord\exp\left[ -i \int_0^{1/2}dt\ H(1-t)\right] = \antitord\exp\left[ i \int_0^{1/2}dt\ \mcs H(t)\mcs^{-1}\right]$ (where $\antitord$ is the antitime ordering symbol), then bring the CS operators outside the anti-time-ordered exponential.  Equation \eqref{eq:identity_TRS} is obtained similarly using Eq.~\eqref{eq:TRS_Hamiltonian_floquet} and the antiunitarity of $\mct$.  Comparing Eq.~\eqref{eq:composition_with_LG} with Eqs.~\eqref{eq:identity_CS} and \eqref{eq:identity_TRS}, we then obtain the two identities \eqref{eq:composition_CS} and \eqref{eq:composition_TRS} by noting
\beq
    U_\text{LG}^{(X)} = \tord\exp\left(-i\int_0^1dt\ \frac{1}{2}H(t/2) \right) = \tord\exp\left(\int_0^{1/2}dt\ H(t) \right),
\eeq
in which the first equality holds by definition.

We can now confirm that composition preserves CS and TRS (if either or both is present in $\mathbb{S}$).  In the case of CS, we use Eq.~\eqref{eq:composition_CS} to obtain
\beq
    \mcs(U\star U_\text{LG})\mcs^{-1} = \mcs^2 U_\text{LG}^{(X)\dagger}\mcs^{-1} U U_\text{LG}^{(X)}\mcs^{-1} =U_\text{LG}^{(X)\dagger}\mcs U U_\text{LG}^{(X)}\mcs^{-1} = U_\text{LG}^{(X)\dagger}U^\dagger \mcs U_\text{LG}^{(X)}\mcs^{-1}= (U\star U_\text{LG})^\dagger,\label{eq:composition_preserves_CS}
\eeq
where we obtain the second equality from Eq.~\eqref{eq:CS_inverse} and the third equality from Eq.~\eqref{eq:CS_unitary}.
Similarly, in the case of TRS, we use Eq.~\eqref{eq:composition_TRS} to obtain
\beq
    \mct(U\star U_\text{LG})\mct^{-1} = \mct^2 U_\text{LG}^{(X)\dagger}\mct^{-1} U U_\text{LG}^{(X)}\mct^{-1} =U_\text{LG}^{(X)\dagger}\mct U U_\text{LG}^{(X)}\mct^{-1} =U_\text{LG}^{(X)\dagger}U^\dagger\mct U_\text{LG}^{(X)}\mct^{-1}= (U\star U_\text{LG})^\dagger,\label{eq:composition_preserves_TRS}
\eeq
where we obtain the second equality by noting that
\beq
    \mct^2 = (\sgn \mct^2 )\mathbb{I} \text{ [hence } \mct^{-1}=(\sgn \mct^2)\mct\text{]},\label{eq:TRS_inverse}
\eeq
and the third equality by using Eq.~\eqref{eq:TRS_unitary}.

Thus, we have shown that for all symmetry classes $\mathbb{S}$, the composition $U\star U_\text{LG}$ is in $\mathbb{S}$.

\subsection{Composition is unique up to homotopy}\label{app:Composition is unique up to homotopy}

In a given symmetry class $\mathbb{S}$, consider two drives, $H_0(t)$ and $H_1(t)$, that each generate the same unitary $U_\text{LG}$.  From Eq.~\eqref{eq:composition_with_LG}, the two corresponding definitions of $U\star U_{\text{LG}}$ (given some unitary $U$ in $\mathbb{S}$) are
\bseq
\begin{align}
    U_0 &\equiv \left(\tord e^{-i\int_{1/2}^1 dt\ H_0(t) } \right) U \left( \tord e^{-i\int_0^{1/2}dt\ H_0(t) }\right),\\
    U_1 &\equiv \left(\tord e^{-i\int_{1/2}^1 dt\ H_1(t) } \right) U \left( \tord e^{-i\int_0^{1/2}dt\ H_1(t) }\right).
\end{align}
\eseq
Our task is to show that $U_0$ and $U_1$ are HE in $\mathbb{S}$.

Suppose $\mathbb{S}$ has CS.  Then, by Eq.~\eqref{eq:composition_CS},
\bseq
\begin{align}
    U_0 &= \mcs U_{\text{LG},0}^{(X)\dagger}\mcs^{-1} U U_{\text{LG},0}^{(X)} ,\\
    U_1 &= \mcs U_{\text{LG},1}^{(X)\dagger}\mcs^{-1} U U_{\text{LG},1}^{(X)} ,
\end{align}
\eseq
where $U_{\text{LG},j}^{(X)}$ is generated by $H_j^{(X)}(t)=\frac{1}{2}H_j(\frac{1}{2}t)$ ($j=0,1$).

Since $U_{\text{LG},0}^{(X)}$ and $U_{\text{LG},1}^{(X)}$ are each HE to the identity, they are HE to each other (all within class $X$).  Let $U_{\text{LG}}^{(X)}(s)$ be a homotopy between them.  Then, the following homotopy connects $U_0$ and $U_1$:
\beq
    U(s) = \mcs U_{\text{LG}}^{(X)\dagger}(s)\mcs^{-1} U U_{\text{LG}}^{(X)}(s),
\eeq
and this homotopy is in $\mathbb{S}$ by construction.

If $\mathbb{S}$ has TRS, we can do the same steps starting from Eq.~\eqref{eq:composition_TRS} instead of Eq.~\eqref{eq:composition_CS}.

\section{Further details on the equivalence of LGE to HE}\label{app:Further details on the equivalence of LGE to HE}

\subsection{Classes without CS or TRS}\label{app:Classes without CS or TRS}
We consider classes A, D, and C, that is, the symmetry classes that at most have PHS.  In the main text, we provided a definition for LGE in these symmetry classes based on the definition of composition as a simple product [Eq.~\eqref{eq:LGE_def_simpler}].  In this section, we show that the alternative approach of defining composition by \eqref{eq:composition_with_LG} [with a resulting change in the definition of LGE given by Eq.~\eqref{eq:LGE_def}] leads to exactly the same classification.  We do this by showing that LGE based on \eqref{eq:composition_with_LG} is equivalent to HE.  Throughout this section, let us use LGE to refer to equivalence based on the definition \eqref{eq:composition_with_LG}.  

To begin, let us show that LGE is equivalent to another condition.  In particular, we show that $U_1$ and $U_0$ are LGE if and only if there are locally generated unitaries $V_1$ and $V_2$, with PHS if present, satisfying
\beq
    U_1 = V_1 U_0 V_2^\dagger.\label{eq:LGE_canonical_relation_ADC}
\eeq
Writing $V_2^\dagger$ here instead of $V_2$ is entirely a matter of convenience; we have made this choice so the calculations involve time-ordered exponentials more often than anti-time-ordered.

Suppose first that  \eqref{eq:LGE_canonical_relation_ADC} holds.  Then, $V_1$ and $V_2$ are generated by some $h_1(t)$ and $h_2(t)$, respectively, that have PHS (if present).  Then we note
\bseq
\begin{align}
    V_1 U_0 V_2^\dagger &= ( \tord e^{-i \int_0^1 dt\ h_1(t)} ) U_0 \antitord e^{i \int_0^1 dt\ h_2(t)} \\
    &= ( \tord e^{-i \int_{1/2}^1 dt\ H(t)} ) U_0 \tord e^{-i \int_0^{1/2} dt\ H(t)},
\end{align}
\eseq
where
\beq
    H(t) = 
    \begin{cases}
        -2 h_2(1-2t) & 0 < t < 1/2,\\
        2 h_1(2t-1) & 1/2 < t < 1.
    \end{cases}
\eeq
From Eq.~\eqref{eq:PHS_Hamiltonian_floquet}, we see that $H(t)$ inherits PHS (if present) from $h_1$ and $h_2$.  Thus, $U_0$ and $U_1$ are LGE.  In an entirely similar way, we may show that $U_0$ and $U_1$ being LGE implies \eqref{eq:LGE_canonical_relation_ADC}; in particular, we define $h_1(t) = \frac{1}{2}H\blp\frac{1}{2}(t+1)\brp$ and $h_2(t)= -\frac{1}{2}H\blp\frac{1}{2}(1-t)\brp$.

Next, we show that HE is equivalent to \eqref{eq:LGE_canonical_relation_ADC}.  Suppose first that  \eqref{eq:LGE_canonical_relation_ADC} holds.  Then, $V_1$ and $V_2$ are generated by some $h_1(t)$ and $h_2(t)$, respectively, that have PHS (if present).  We define $V_1(s)$ and $V_2(s)$ to be generated by $h_1(t)$ and $h_2(t)$, respectively, from $t=0$ to $s$.  Then, we define
\beq
    U(s) = V_1(s) U_0 V_2^\dagger(s),
\eeq
which satisfies $U(0) =U_0$ and $U(1) = U_1$ by construction.  Furthermore, if PHS is present, then $V_1(s)$ and $V_2(s)$ each have it and hence $U(s)$ does also.  Thus, we have shown that the condition \eqref{eq:LGE_canonical_relation_ADC} implies that $U_1$ and $U_0$ are HE.

The main step is showing that HE implies the condition \eqref{eq:LGE_canonical_relation_ADC}.  Given a homotopy $U(s)$ from $U_0$ to $U_1$, we define two Hamiltonians that each have PHS (if present):
\bseq
\begin{align}
    h_1(s) &= i\frac{1}{2}\frac{\pd U(s)}{\pd s} U^\dagger(s),\\
    h_2(s) &= -i\frac{1}{2}U^\dagger(s)\frac{\pd U(s)}{\pd s}.
\end{align}
\eseq
We readily verify that $i\frac{1}{2}\pd_s U(s) = h_1(s)U(s) = - U(s)h_2(s)$.  Adding these equations, we thus obtain
\beq
    i\frac{\pd U(s)}{\pd s} = h_1(s) U(s)- U(s)h_2(s).
\eeq
As in the main text, we have arrived at a first-order differential equation, and we proceed to write its unique solution in a convenient form:
\beq
    U(s) = V_1(s) U_0 V_2^\dagger(s),
\eeq
where $V_1(s)$ and $V_2(s)$ are generated by $h_1(t)$ and $h_2(t)$, respectively, from $t=0$ to $s$.  Note that for all $s$, $V_1(s)$ and $V_2(s)$ have PHS, if present.  Setting $s=1$, we obtain the condition \eqref{eq:LGE_canonical_relation_ADC}, which completes the proof.

\subsection{Classes with CS}\label{app:Classes with CS}
We show that LGE is equivalent to the condition \eqref{eq:LG_canonical_relation_chiral} in the symmetry classes $\mathbb{S}$ with CS.  Then, we provide some further details for the proof (from Sec. \ref{sec:Proof that LGE and HE are equivalent}) that \eqref{eq:LG_canonical_relation_chiral} is equivalent to HE.

Consider the symmetry classes with CS.  If two unitaries $U_0$ and $U_1$ are LGE, then by definition we have $U_1 = U_0 \star U_\text{LG}$ for some $U_\text{LG}$ that is locally generated in $\mathbb{S}$, and then Eq.~\eqref{eq:LG_canonical_relation_chiral} holds by the identity \eqref{eq:composition_CS}.

Conversely, suppose that $U_0$ and $U_1$ are related by Eq.~\eqref{eq:LG_canonical_relation_chiral} for some $U_\text{LG}^{(X)}$ that is locally generated in $X$.  By assumption, there is some $H^{(X)}(t)$ in $X$ that generates $U_\text{LG}^{(X)}$.  We then define
\beq
    H(t) =
    \begin{cases}
        2H^{(X)}(2t) & 0 \le t < 1/2,\\
        -2 \mcs H^{(X)}(2-2t)\mcs^{-1} & 1/2 < t \le 1.
    \end{cases}\label{eq:H(t)_CS}
\eeq
There is no need to define $H(t)$ at the specific point $t=1/2$, since this point is a set of measure zero in the integration that appears in the time-ordered exponential.

We claim that $H(t)$ is in $\mathbb{S}$.  To see this, first recall that $X$ has PHS if and only if $\mathbb{S}$ does.  It is clear from Eq.~\eqref{eq:PHS_Hamiltonian_floquet} that $H(t)$ inherits PHS from $H^{(X)}(t)$ if present.  It is also straightforward to verify, using Eq.~\eqref{eq:CS_inverse}, that $H(t)$ has CS [Eq.~\eqref{eq:CS_Hamiltonian_floquet}].  TRS, if present, follows immediately because it is a combination of CS and PHS.

We then define $U_\text{LG}$ to be generated by $H(t)$.  By construction, we have $H^{(X)}(t) = \frac{1}{2}H(\frac{1}{2} t)$ for $0\le t\le 1$.  Thus, by the identity \eqref{eq:composition_CS}, we have $U_0\star U_\text{LG} = \mcs U_\text{LG}^{(X)\dagger}\mcs^{-1}U_0U_\text{LG}^{(X)}$, which shows that $U_0$ and $U_1$ are LGE.  This completes the proof that LGE is equivalent to the condition \eqref{eq:LG_canonical_relation_chiral} in the symmetry classes with CS.

Let us make one comment on this proof.  If we take \eqref{eq:LG_canonical_relation_chiral} as given, we have some $H^{(X)}(t)$ that generates $U_\text{LG}^{(X)}$ (by assumption), and we construct $U_\text{LG}$ as being generated by $H(t)$ [defined by \eqref{eq:H(t)_CS}].  If we then take this $U_\text{LG}$ as the input to our proof that LGE implies \eqref{eq:LG_canonical_relation_chiral}, the $H^{(X)}(t)$ that we construct will be identical to the original $H^{(X)}(t)$ if we choose to generate $U_\text{LG}$ with $H(t)$.  However, this agreement is not essential to the proof; we could, in principle, generate $U_\text{LG}$ by some $H'(t)$ that differs from $H(t)$, and then the quantity $H^{'(X)}(t)=\frac{1}{2}H'(\frac{1}{2}t)$ that is constructed in the proof would differ from $H^{(X)}(t)$.

Next, we provide some further comments on the proof that \eqref{eq:LG_canonical_relation_chiral} is equivalent to HE.

We can confirm that $U_\text{LG}^{'(X)}$ is generated by $-\mcs h^{(X)}(1-t) \mcs^{-1}$ as follows:
\bseq
\begin{align}
    U_\text{LG}^{'(X)}&=  \mcs \left\{ \tord\exp[-i\int_0^1 dt\ h^{(X)}(t)]\right\}^\dagger\mcs^{-1} \label{eq:ULGprime CS eq1}\\
    &= \mcs \tord \exp[i\int_0^1 dt\  h^{(X)}(1-t)] \mcs^{-1}\label{eq:ULGprime CS eq2}, 
\end{align}
\eseq
where the first equality holds by definition and the second by relabeling $t\to 1-t$.  Then, we only need to bring the CS operators inside the time-ordered exponential.

In the proof that \eqref{eq:LG_canonical_relation_chiral} is equivalent to HE, it is important to note that the $U_\text{LG}^{(X)}$ in \eqref{eq:LG_canonical_relation_chiral} need not be unique.  To make this point more explicit, consider taking \eqref{eq:LG_canonical_relation_chiral} as given; then, we have some $H^{(X)}(t)$ that (by assumption) generates $U_\text{LG}^{(X)}$, and we use this $H^{(X)}(t)$ to construct the homotopy \eqref{eq:homotopy_for_alternate_implies_HE}.  If we take this homotopy as the input to our proof that HE implies \eqref{eq:LG_canonical_relation_chiral}, we obtain $U_\text{LG}^{'(X)}$ generated by $H^{'(X)}(t)= - \mcs h^{(X)}(1-t)\mcs^{-1}= -\frac{1}{2}i (\pd_t U^\dagger)(1-t)U(1-t)$, which seem to differ from $U_\text{LG}^{(X)}$ and $H^{(X)}(t)$.  However, for our purpose of showing that LGE and HE are equivalent, the $U_\text{LG}^{(X)}$ that appears in \eqref{eq:LG_canonical_relation_chiral} need not be unique, and even if it is unique, the Hamiltonian that generates it need not be unique.

\subsection{Classes with TRS}\label{app:Classes with TRS}
We start by showing that LGE is equivalent to the condition \eqref{eq:LG_canonical_relation_TRS} in the symmetry classes $\mathbb{S}$ with TRS.  LGE (of two unitaries $U_0$ and $U_1$) immediately implies Eq.~\eqref{eq:LG_canonical_relation_TRS} due to the identity \eqref{eq:composition_TRS}.  Conversely, if we instead assume Eq.~\eqref{eq:LG_canonical_relation_TRS}, then we define
\beq
    H(t) =
    \begin{cases}
        2H^{(X)}(2t) & 0 \le t < 1/2,\\
        -2 \mct H^{(X)}(2-2t)\mct^{-1} & 1/2 < t \le 1.
    \end{cases}\label{eq:H(t)_TRS}
\eeq
As in the CS case [Appendix \ref{app:Classes with CS}], it is straightforward to verify that $H(t)$ is in $\mathbb{S}$ [now using Eq.~\eqref{eq:TRS_inverse}].  Then we again define $U_\text{LG}$ to be generated by $H(t)$, and we obtain $U_0\star U_\text{LG} = \mct U_\text{LG}^{(X)\dagger}\mct^{-1} U_0 U_\text{LG}^{(X)}$ using the identity \eqref{eq:composition_TRS}.

We proceed to show that HE (in a class $\mathbb{S}$ with TRS) is equivalent to the condition \eqref{eq:LG_canonical_relation_TRS}, following the same approach as in Sec. \ref{sec:Proof that LGE and HE are equivalent}.  One direction is trivial: given \eqref{eq:LG_canonical_relation_TRS}, we follow exactly the same steps as in the paragraph above Eq.~\eqref{eq:homotopy_for_alternate_implies_HE}, now using the $U_\text{LG}^{(X)}$ that appears in \eqref{eq:LG_canonical_relation_TRS}; then \eqref{eq:homotopy_for_alternate_implies_HE} (with $\mcs$ replaced by $\mct$) is the required homotopy.  We may readily verify that $U(s)$ has PHS (if present).  Also, $U(s)$ has TRS, as we can see from a calculation very similar to \eqref{eq:composition_preserves_TRS}.  Thus, $U(s)$ is in the given symmetry class $\mathbb{S}$.

It remains to show the other direction: that HE implies \eqref{eq:LG_canonical_relation_TRS}.  Given a homotopy $U(s)$ that connects $U_0$ to $U_1$, we define $h^{(X)}(s)$ by Eq.~\eqref{eq:HX(s)}, where the class $X$ is given by Eq.~\eqref{eq:X}.  The same argument as given below Eq.~\eqref{eq:HX(s)} shows that $h^{(X)}(s)$ ($0\le s \le 1$) is a local Hamiltonian in class $X$.

By construction, we have $i \pd_s U(s)= 2i h^{(X)}(s) U(s)$.  We also have $i \pd_s U(s) = 2U(s)\mct h^{(X)}(s) \mct^{-1}$; to see this, we note that $\mct h^{(X)}(s) \mct^{-1}= -2i [\pd_s U^\dagger(s)] U(s)= 2i U^\dagger(s) \pd_s U(s)$, where the first equality holds by the TRS of $U(s)$ (and the homotopy being sufficiently nice to allow commuting $S$ with the derivative) and the second equality holds because $U^\dagger(s)U(s) = \mathbb{I}$. Adding these equations yields
\beq
    i \frac{\pd U(s)}{\pd s} =   h^{(X)}(s)U(s) + U(s)\mct h^{(X)}(s)\mct^{-1}.\label{eq:diff_eqn_TRS}
\eeq
The key point is that Eq.~\eqref{eq:diff_eqn_TRS}, being a first-order differential equation, has a unique solution given the initial condition $U(0)= U_0$.  We can therefore obtain a convenient formal solution by making an ansatz and checking that it works.

Our ansatz is
\beq
    U(s) = u_\text{LG}^{(X)}(s) U_0 \mct u_\text{LG}^{(X)\dagger}(s) \mct^{-1},\label{eq:ansatz_TRS}
\eeq
where $u_\text{LG}^{(X)}(s)$ is generated by $h^{(X)}(t)$ from $t=0$ to $s$.  Then $u_\text{LG}^{(X)}(0)=\mathbb{I}$, so $U(0)= U_0$, as required.  The differential equation \eqref{eq:diff_eqn_TRS} holds due to $i\pd_s u_\text{LG}^{(X)}= h^{(X)}(s)u_\text{LG}^{(X)}(s)$ (and the adjoint of this equation); c.f. the antiunitarity of $\mct$.  Thus, Eq.~\eqref{eq:ansatz_TRS} is a valid expression for the given homotopy $U(s)$.

We have therefore shown
\beq
    U_1 = u_\text{LG}^{(X)} U_0 \mct u_\text{LG}^{(X)\dagger}\mct^{-1},\label{eq:U1_alternate_TRS}
\eeq
where $u_\text{LG}^{(X)}\equiv u_\text{LG}^{(X)}(1)$ is locally generated in class $X$.  As we now show, this form for $U_1$ is the same as \eqref{eq:LG_canonical_relation_TRS} up to trivial transformations.  We define $U_\text{LG}^{'(X)} = \mct u_\text{LG}^{(X)\dagger}\mct^{-1}$; then $U_1 = \mct^{-1} U_\text{LG}^{'(X)\dagger} \mct U_0  U_\text{LG}^{(X)}$, which establishes Eq.~\eqref{eq:LG_canonical_relation_TRS} [c.f. Eq.~\eqref{eq:TRS_inverse}].  Note that $U_\text{LG}^{'(X)}$ is generated by $H^{'(X)}(t)\equiv\mct h^{(X)}(1-t) \mct^{-1}$ and is thus locally generated in class $X$. We can see this from Eqs.~\eqref{eq:ULGprime CS eq1} and \eqref{eq:ULGprime CS eq2} with $\mct$ replacing $\mcs$ (note that the sign inside the exponent will flip sign in this case once we bring the TRS operators past the $i$).  This completes the proof of the equivalence of HE to \eqref{eq:LG_canonical_relation_TRS}, and thus also of HE to LGE in the classes with TRS.

\section{Classification with restricted number of orbitals}\label{app:Classification with restricted number of orbitals}
In this Appendix, we show that restricting the number of generalized orbitals to be even or to be a multiple of four [according to \eqref{eq:alpha conditions}] has no effect on the classification of Hamiltonians.  Although we only need to consider the symmetry classes A, AI, and AII, the argument is more clearly presented by considering a general symmetry class.

Write $\mathcal{H}^\mathbb{S}$ for the set of gapped, local Hamiltonians in symmetry class $\mathbb{S}'$, and write $\mathcal{H}_r^{\mathbb{S}'}$ for the subset for which the number of generalized orbitals is restricted to be a multiple of four (later we also treat the case of restriction to an even number).  We have already discussed the notions of homotopy equivalence ($\approx)$ and stable homotopy equivalence ($\sim$) on $\mathcal{H}^{\mathbb{S}'}$ in Sec. \ref{sec:hermitian_equivalence}.  We define homotopy equivalence ($\overset{r}{\approx}$) and stable homotopy equivalence ($\overset{r}{\sim}$) on $\mathcal{H}_r^{\mathbb{S}'}$ in the same way with the additional requirements that the homotopy $H(s)$ and the trivial Hamiltonians $H_0^{(0)}$ and $H_1^{(0)}$ [in Eq.~\eqref{eq:stable_homotopy_Hamiltonians}] must be in $\mathcal{H}_r^{\mathbb{S}'}$.

Our task is to show that two Hamiltonians in the restricted space $\mathcal{H}_r^{\mathbb{S}'}$ are equivalent up to stable homotopy \emph{within} $\mathcal{H}_r^{\mathbb{S}'}$ if and only if they are equivalent up to stable homotopy within the larger set $\mathcal{H}^{\mathbb{S}'}$.  That is, we must show
\beq
H_0\overset{r}{\sim}H_1\text{ if and only if }H_0 \sim H_1.\label{eq:equivalence for multiples of 4}
\eeq
This equivalence implies that the Hamiltonians in $\mathcal{H}_r^{\mathbb{S}'}$ have the same classification as the Hamiltonians in $\mathcal{H}^{\mathbb{S}'}$.

One direction of the equivalence is immediate: If $H_0 \overset{r}{\sim}H_1$, then $H_0 \sim H_1$, because the homotopy that connects $H_0\oplus H_0^{(0)}$ and $H_1\oplus H_1^{(0)}$ within $\mathcal{H}_r^{\mathbb{S}'}$ is also a homotopy within $\mathcal{H}^{\mathbb{S}'}$ (since $\mathcal{H}_r^{\mathbb{S}'}$ is a subset of $\mathcal{H}^{\mathbb{S}'}$).

To show the other direction, consider $H_0$ and $H_1$ in $\mathcal{H}_r^{\mathbb{S}'}$ with $H_0\overset{r}\sim H_1$.  The number of generalized orbitals for $H_0$ and $H_1$ is $4n_0$ and $4n_1$, respectively (for some integers $n_0,n_1$).  By assumption, there are trivial Hamiltonians $H_0^{(0)}$ and $H_1^{(0)}$ (with $\alpha_0$ and $\alpha_1$ generalized orbitals, respectively) for which $H_0 \oplus H_0^{(0)} \approx H_1\oplus H_1^{(0)}$.  Then there is a homotopy $H(s)$, with $4n_0 + \alpha_0=4n_1+\alpha_1$ generalized orbitals, from $H_0 \oplus H_0^{(0)}$ to $H_1\oplus H_1^{(0)}$.

If $\alpha_0$ is a multiple of four, then so is $\alpha_1$, and $H(s)$ is then the required homotopy within $\mathcal{H}_r^{\mathbb{S}'}$.  Otherwise, let $\alpha'\ge1$ be any integer for which $\alpha_0+\alpha'$ is a multiple of four, and let $H'$ be a trivial Hamiltonian in $\mathbb{S}'$ with $\alpha'$ generalized orbitals.  (For $\mathbb{S}'=$AII, the number of generalized orbitals is always even, so $\alpha'=2$ suffices.)  Then $H_0^{(0)}\oplus H'$ and $H_1^{(0)}\oplus H'$ are trivial Hamiltonians in $\mathcal{H}_r^\mathbb{S}$, and we can define the following homotopy from $H_0 \oplus H_0^{(0)}\oplus H'$ to $H_1 \oplus H_1^{(0)}\oplus H'$:
\beq
    H_r(s) = H(s) \oplus H',\label{eq:Hr(s)}
\eeq
which has $4n_0+\alpha_0+\alpha'$ generalized orbitals and thus is in $\mathcal{H}_r^{\mathbb{S}'}$.  This homotopy implies $H_0\overset{r}{\sim}H_1$, which completes the proof of \eqref{eq:equivalence for multiples of 4}.

We can also define $\mathcal{H}_r^{\mathbb{S}'}$ to be the subset of $\mathcal{H}_r^{\mathbb{S}'}$ with the number of generalized orbitals defined to be even.  Then, repeating the same argument with $\alpha_0+\alpha'$ only required to be even, we may show the same equivalence \eqref{eq:equivalence for multiples of 4} for the case in which the Hamiltonians are restricted to have an even number of orbitals rather than a multiple of four.  Thus, we conclude that the restrictions \eqref{eq:alpha conditions} that we made to define an inverse to $H_U$ do not affect the topological classification. 

Finally, we note that, just as in the case of $\mathcal{H}^{\mathbb{S}'}$, there is no effect on the classification if we restrict our attention to the subset of \emph{flattened} Hamiltonians within $\mathcal{H}_r^{\mathbb{S}'}$.  It is on this subset that the inverse to $H_U$ is defined.

\section{Mapping between unitaries and Hamiltonians preserves stable homotopy\label{app:Mapping between unitaries and Hamiltonians preserves stable homotopy}}
In this Appendix, we verify that, in each symmetry class, the one-to-one mapping $U\mapsto H_U$ preserves equivalence up to stable homotopy [that is, we verify \eqref{eq:mapping preserves stable homotopy}].  In particular, here we consider two unitaries that are in the same symmetry class $\mathbb{S}$, but that may have different numbers of generalized orbitals.  Equivalently, since the mapping is one-to-one, we can consider two Hamiltonians in the same symmetry class $\mathbb{S}'$ (which is related to $\mathbb{S}$ according to Table \ref{tab:symmetry_transformations}), possibly with different numbers of generalized orbitals.  In the cases $\mathbb{S}'=$A, AI, or AII, we understand each instance of $\sim$ and $\approx$ below to in fact refer to the corresponding equivalences $\overset{r}{\sim}$ and $\overset{r}{\approx}$ within the restricted space $\mathcal{H}_r^{\mathcal{S}'}$ that has the condition \eqref{eq:alpha conditions} imposed on the number of generalized orbitals [see Appendix \ref{app:Classification with restricted number of orbitals}].

There are two cases to consider: the five symmetry classes $\mathbb{S}$ that do not have CS (equivalently, the five $\mathbb{S}'$ that do), and the five symmetry classes $\mathbb{S}$ that have CS (equivalently, the five $\mathbb{S}'$ that do not).  In each case, we first verify that
the mapping preserves HE, that is,
\beq
    U_0 \approx U_1 \iff H_{U_0}\approx H_{U_1},\label{eq:mapping preserves HE}
\eeq
in which $U_0$ and $U_1$ have the same number of generalized orbitals (or, equivalently, $H_{U_0}$ and $H_{U_1}$ have the same number, which is greater by a factor of two in the case that the unitaries do not have CS).  We then use \eqref{eq:mapping preserves HE} to obtain the desired result \eqref{eq:mapping preserves stable homotopy}.

If $\mathbb{S}$ is one of the five classes without CS, then the mapping is given by Eq.~\eqref{eq:H_U_def}.  Suppose $U_0\approx U_1$.  Then we may use the homotopy $U(s)$ between $U_0$ and $U_1$, and the mapping~\eqref{eq:H_U_def}, to define a family of local Hamiltonians
\beq
    H(s) = 
    \bpmat
        0 & U(s) \\
        U^\dagger(s) & 0
    \epmat,\label{eq:homotopy_nochiral}
\eeq
which, by construction, is a homotopy between $U_{H_0}$ and $U_{H_1}$.  From Sec. \ref{sec:Unitaries without chiral symmetry}, we know that $H(s)$ has the appropriate symmetries [note that $U(s)$ by assumption has the same symmetries as $U_0$ and $U_1$].  Thus, $H_{U_0}\approx H_{U_1}$.  Conversely, given two flattened, local, HE Hamiltonians in a symmetry class $\mathcal{S}'$ with CS, we have unique $U_0$, $U_1$ for which the two Hamiltonians are equal to $H_{U_0}$ and $H_{U_1}$.  The given homotopy $H(s)$ between $H_{U_0}$ and $H_{U_1}$ then maps (under the inverse mapping $H\mapsto H_U$) to a homotopy $U(s)$ given by \eqref{eq:homotopy_nochiral}, and, as shown in Sec. \ref{sec:Unitaries without chiral symmetry}, $U(s)$ is local and has the appropriate symmetries.  Thus, $U_0\approx U_1$, and so \eqref{eq:mapping preserves HE} is obtained for all $\mathbb{S}$ without CS.

If $\mathbb{S}$ is one of the five classes with CS, then the mapping is given by Eq.~\eqref{eq:HU_def_chiral}.  Similar to the argument in the previous paragraph, we define
\beq
    H(s) = S U(s),\label{eq:homotopy_chiral}
\eeq
where we either take (a) $U(s)$ to be a homotopy between $U_0$ and $U_1$ (supposing we are given $U_0\approx U_1$), or (b) $H(s)$ to be a given homotopy between $H_{U_0}$ and $H_{U_1}$ (supposing we are given $H_{U_0}\approx H_{U_1}$).  In case (b), the homotopy $U(s)$ is defined by the inverse mapping, i.e., $U(s)= S H(s)$.  We thus obtain \eqref{eq:mapping preserves HE} in the remaining five symmetry classes.

One immediate consequence of \eqref{eq:mapping preserves HE} is that a unitary is trivial (i.e., HE to the identity) if and only if the corresponding Hamiltonian is trivial.  To obtain the desired result \eqref{eq:mapping preserves stable homotopy}, it then suffices to show
\beq
    H_{U_0\oplus U_1} = H_{U_0} \oplus H_{U_1},\label{eq:additivity of HU}
\eeq
in which $U_0$ and $U_1$ are arbitrary local unitaries in the same symmetry class.  (We only need the special case in which one of these unitaries is trivial, but the general case is no more difficult.)  Equation \eqref{eq:additivity of HU} follows immediately from the definitions \eqref{eq:H_U_def} and \eqref{eq:HU_def_chiral}, because we indeed have
\bseq
\begin{align}
    \bpmat
        0 & U_0 \oplus U_1\\
        (U_0\oplus U_1)^\dagger & 0 
    \epmat
    &=
    \bpmat
        0 & U_0 \\
        U_0^\dagger & 0 
    \epmat
    \oplus
    \bpmat
        0 & U_1 \\
        U_1^\dagger & 0 
    \epmat,\\
    S(U_0\oplus U_0) &= (S U_1)\oplus (SU_1).
\end{align}
\eseq

From \eqref{eq:mapping preserves HE} and \eqref{eq:additivity of HU}, which we have shown in all symmetry classes, we immediately obtain
\beq
    U_0\oplus U_0^{(0)} \approx U_1 \oplus U_1^{(0)} \iff H_{U_0\oplus U_0^{(0)} } \approx H_{U_1\oplus U_1^{(0)}} \iff H_{U_0}\oplus H_{U_0^{(0)}} \approx H_{U_1}\oplus H_{U_1^{(0)}},
\eeq
which demonstrates \eqref{eq:mapping preserves stable homotopy} in all symmetry classes.

\section{Further details on the mapping from non-chirally-symmetric unitaries to Hamiltonians}\label{app:Further details on the mapping from non-chirally-symmetric unitaries to Hamiltonians}
Here we collect some routine calculations omitted from the main text.

First, we verify that the right-hand side of Eq.~\eqref{eq:H_U_def} is the most general form for a flattened, chirally symmetric Hamiltonian.  Consider a Hamiltonian $H$ with CS.  Note that $H$ necessarily has an even number of orbital degrees of freedom, and there is some sublattice basis in which CS acts as $\sigma^z$.  We write $H$ in this basis as
\beq
H=
\bpmat
a & b\\
c& d
\epmat.
\eeq
We then note
\beq
    \sigma^z
\bpmat
a & b\\
c& d
\epmat
\sigma^z =
\bpmat
    a & -b \\
    -c & d
\epmat,
\eeq
hence, the CS condition, Eq.~\eqref{eq:CS_ham}, implies $a=d =0$.  Since $H$ is Hermitian, $c= b^\dagger$.  Finally, since $H$ is flattened, $b$ is unitary.

Next, we prove the equivalences of symmetry actions shown in the main text.  To prove \eqref{eq:ThetaPHS_equivalences}, we note
\bequ
    (\mathbb{I}_2\otimes \Theta) H_U (\mathbb{I}_2\otimes \Theta)^{-1} = 
    \bpmat
        0 & \Theta U\Theta^{-1}\\
        \Theta U^\dagger \Theta^{-1} & 0 
    \epmat,
\eequ
which implies that the left-hand side equals $H_U$ if and only if $\Theta U\Theta^{-1} = U$.  This statement is exactly the first line of \eqref{eq:ThetaPHS_equivalences}; the second line may be obtained either by a similar calculation or by recalling that $\sigma_z\otimes\mathbb{I}$ is a CS for $H_U$.

Similarly, \eqref{eq:ThetaTRS_equivalences} is proven by noting
\bequ
    (\sigma_x\otimes\Theta) H_U (\sigma_x\otimes\Theta)^{-1} = 
    \bpmat
        0 & \Theta U^\dagger \Theta^{-1}\\
        \Theta U \Theta^{-1} & 0 
    \epmat,\label{eq:sigmax HU identity}
\eequ
which implies that the left-hand side equals $H_U$ if and only if $\Theta U\Theta^{-1} = U^\dagger$.  This statement is exactly the first line of \eqref{eq:ThetaTRS_equivalences}; the second line may be obtained either by a similar calculation or by recalling that $\sigma_z\otimes\mathbb{I}$ is a CS for $H_U$.

\section{Canonical forms for symmetry operators}\label{app:Canonical forms for symmetry operators}
In this Appendix, we present the basis transformations that bring the symmetry operators (in the four symmetry classes with both TRS and PHS) from their standard forms to the forms we used in the main text.

TRS and PHS can be written as 
\beq
    \mct = \mathcal{K}U_\mct,\ \mcc=\mathcal{K}U_\mcc,\label{eq:TRS and PHS in terms of unitaries}
\eeq
where $\mathcal{K}$ is complex conjugation and $U_\mct$ and $U_\mcc$ are unitary.  From Eq.~\eqref{eq:CS_as_CT}, the CS operator is then given by
\beq
    \mcs = e^{i\delta} U_\mcc^* U_\mct.\label{eq:CS in terms of unitaries}
\eeq

Complex conjugation is defined relative to a certain basis.  Under a change of basis by a unitary matrix $V$, the unitary matrices $U_\mct$ and $U_\mcc$ transform as
\beq
    U_\mct\to V^* U_\mct V^\dagger,\ U_\mcc\to V^* U_\mcc V^\dagger,
\eeq
where the star appears due to $\mathcal{K}$.

In Table \ref{tab:Table of basis transformations}, we present the changes of basis needed to obtain the canonical forms used in the main text.  The starting forms are taken from Table I of Ref.~\cite{RoyPeriodic2017}.  In three of the symmetry classes, two different basis transformations are presented because different forms are needed for Hamiltonians and for unitaries.  From Eq.~\eqref{eq:CS in terms of unitaries}, it may be verified that the CS operator in the new basis is $\mcs = \sigma_z\otimes\id$ (up to a phase) in each case.

\begin{table}[htp]
    \caption{\label{tab:Table of basis transformations}
    Basis transformations for TRS and PHS in the four symmetry classes where both are present.}
    \begin{ruledtabular}
    \begin{tabular}{ l C C C C C}
    & \multicolumn{2}{c}{Starting forms} & \text{Change of basis}& \multicolumn{2}{c}{New forms} \\
    \cmidrule(lr){2-3}\cmidrule(lr){5-6}
    \text{AZ class} &
    U_\mct & U_\mcc & V & V^* U_\mct V^\dagger & V^* U_\mcc V^\dagger\\
    \colrule
    BDI & \id & \sigma_x\otimes \id & \frac{1}{\sqrt{2}}\bpmat 1&0&1&0\\0&1&0&1\\-1&0&1&0\\0&-1&0&1  \epmat &\id & \sigma_z\otimes\id\\
    \\
     &  &  & \frac{1}{\sqrt{2}}\bpmat 0&1&0&1\\1&0&1&0\\-i&0&i&0\\0&-i&0&i  \epmat &\sigma_z\otimes\id & \id\\
    DIII & \id\otimes i\sigma_y & \sigma_x\otimes \id &\frac{1}{\sqrt{2}}\bpmat i&0&0&1\\0&-i&1&0\\-i&0&0&1\\0&i&1&0  \epmat  &-\sigma_x\otimes i\sigma_y & \sigma_y\otimes i\sigma_y\\
    \\
    & & &\frac{1}{\sqrt{2}}\bpmat 0&1&i&0\\1&0&0&-i\\-i&0&0&1\\0&i&1&0  \epmat  &\sigma_y\otimes \id & \sigma_x\otimes \id\\
    CII & \id\otimes i\sigma_y & i\sigma_y\otimes\id &\frac{1}{\sqrt{2}}\bpmat 1&0&0&1\\0&-1&1&0\\-1&0&0&1\\0&1&1&0  \epmat &-\id\otimes i\sigma_y & \sigma_z\otimes i\sigma_y\\
    \\
    &     & & \frac{1}{\sqrt{2}}\bpmat 0&-i&i&0\\i&0&0&i\\-i&0&0&i\\0&i&i&0  \epmat & -\sigma_z\otimes i\sigma_y&  \id\otimes i\sigma_y  \\
    CI & \id & i\sigma_y\otimes\id &\frac{1}{\sqrt{2}}\bpmat 0&i&0&1\\i&0&1&0\\0&-i&0&1\\-i&0&1&0  \epmat &\sigma_x\otimes\id & i\sigma_y\otimes\id
    
    \end{tabular}
    \end{ruledtabular}
\end{table}

The new forms from Table \ref{tab:Table of basis transformations}, together with Eq.~\eqref{eq:TRS and PHS in terms of unitaries} and some trivial phase re-definitions, yield the canonical forms listed in Table \ref{tab:Table of canonical forms}, which are used in the main text.
\begin{table}[htp]
    \caption{\label{tab:Table of canonical forms}
    Canonical forms for TRS and PHS in the four symmetry classes where both are present.}
    \begin{ruledtabular}
    \begin{tabular}{ l C C C C}
    & \multicolumn{2}{c}{Unitaries} & \multicolumn{2}{c}{Hamiltonians} \\
    \cmidrule(lr){2-3}\cmidrule(lr){4-5}
    \text{AZ class} &
    \mct & \mcc & \mct & \mcc \\
    \colrule
    BDI & \mathcal{K}\sigma_z\otimes\id & \mathcal{K}  & \mathcal{K} & \mathcal{K}\sigma_z\otimes \id\\
    DIII & \mathcal{K}\sigma_y\otimes \id & \mathcal{K}\sigma_x\otimes \id &\mathcal{K}\sigma_x\otimes \sigma_y & \mathcal{K}\sigma_y\otimes\sigma_y\\
    CII & \mathcal{K}\sigma_z\otimes\sigma_y & \mathcal{K}\id\otimes\sigma_y  &\mathcal{K}\id\otimes \sigma_y & \mathcal{K}\sigma_z\otimes\sigma_y\\
    CI & \mathcal{K}\sigma_x & \mathcal{K}\sigma_y  &\mathcal{K}\sigma_x & \mathcal{K} \sigma_y
    
    \end{tabular}
    \end{ruledtabular}
\end{table}

\end{widetext}

\bibliography{ms}

\begin{thebibliography}{38}%
\makeatletter
\providecommand \@ifxundefined [1]{%
 \@ifx{#1\undefined}
}%
\providecommand \@ifnum [1]{%
 \ifnum #1\expandafter \@firstoftwo
 \else \expandafter \@secondoftwo
 \fi
}%
\providecommand \@ifx [1]{%
 \ifx #1\expandafter \@firstoftwo
 \else \expandafter \@secondoftwo
 \fi
}%
\providecommand \natexlab [1]{#1}%
\providecommand \enquote  [1]{``#1''}%
\providecommand \bibnamefont  [1]{#1}%
\providecommand \bibfnamefont [1]{#1}%
\providecommand \citenamefont [1]{#1}%
\providecommand \href@noop [0]{\@secondoftwo}%
\providecommand \href [0]{\begingroup \@sanitize@url \@href}%
\providecommand \@href[1]{\@@startlink{#1}\@@href}%
\providecommand \@@href[1]{\endgroup#1\@@endlink}%
\providecommand \@sanitize@url [0]{\catcode `\\12\catcode `\$12\catcode `\&12\catcode `\#12\catcode `\^12\catcode `\_12\catcode `\%12\relax}%
\providecommand \@@startlink[1]{}%
\providecommand \@@endlink[0]{}%
\providecommand \url  [0]{\begingroup\@sanitize@url \@url }%
\providecommand \@url [1]{\endgroup\@href {#1}{\urlprefix }}%
\providecommand \urlprefix  [0]{URL }%
\providecommand \Eprint [0]{\href }%
\providecommand \doibase [0]{https://doi.org/}%
\providecommand \selectlanguage [0]{\@gobble}%
\providecommand \bibinfo  [0]{\@secondoftwo}%
\providecommand \bibfield  [0]{\@secondoftwo}%
\providecommand \translation [1]{[#1]}%
\providecommand \BibitemOpen [0]{}%
\providecommand \bibitemStop [0]{}%
\providecommand \bibitemNoStop [0]{.\EOS\space}%
\providecommand \EOS [0]{\spacefactor3000\relax}%
\providecommand \BibitemShut  [1]{\csname bibitem#1\endcsname}%
\let\auto@bib@innerbib\@empty
\bibitem [{\citenamefont {Hasan}\ and\ \citenamefont {Kane}(2010)}]{HasanColloquium2010}%
  \BibitemOpen
  \bibfield  {author} {\bibinfo {author} {\bibfnamefont {M.~Z.}\ \bibnamefont {Hasan}}\ and\ \bibinfo {author} {\bibfnamefont {C.~L.}\ \bibnamefont {Kane}},\ }\bibfield  {title} {\bibinfo {title} {Colloquium: {{Topological}} insulators},\ }\href {https://doi.org/10.1103/RevModPhys.82.3045} {\bibfield  {journal} {\bibinfo  {journal} {Reviews of Modern Physics}\ }\textbf {\bibinfo {volume} {82}},\ \bibinfo {pages} {3045} (\bibinfo {year} {2010})}\BibitemShut {NoStop}%
\bibitem [{\citenamefont {Chiu}\ \emph {et~al.}(2016)\citenamefont {Chiu}, \citenamefont {Teo}, \citenamefont {Schnyder},\ and\ \citenamefont {Ryu}}]{ChiuClassification2016}%
  \BibitemOpen
  \bibfield  {author} {\bibinfo {author} {\bibfnamefont {C.-K.}\ \bibnamefont {Chiu}}, \bibinfo {author} {\bibfnamefont {J.~C.~Y.}\ \bibnamefont {Teo}}, \bibinfo {author} {\bibfnamefont {A.~P.}\ \bibnamefont {Schnyder}},\ and\ \bibinfo {author} {\bibfnamefont {S.}~\bibnamefont {Ryu}},\ }\bibfield  {title} {\bibinfo {title} {Classification of topological quantum matter with symmetries},\ }\href {https://doi.org/10.1103/RevModPhys.88.035005} {\bibfield  {journal} {\bibinfo  {journal} {Reviews of Modern Physics}\ }\textbf {\bibinfo {volume} {88}},\ \bibinfo {pages} {035005} (\bibinfo {year} {2016})}\BibitemShut {NoStop}%
\bibitem [{\citenamefont {Chen}\ \emph {et~al.}(2010)\citenamefont {Chen}, \citenamefont {Gu},\ and\ \citenamefont {Wen}}]{ChenLocal2010}%
  \BibitemOpen
  \bibfield  {author} {\bibinfo {author} {\bibfnamefont {X.}~\bibnamefont {Chen}}, \bibinfo {author} {\bibfnamefont {Z.-C.}\ \bibnamefont {Gu}},\ and\ \bibinfo {author} {\bibfnamefont {X.-G.}\ \bibnamefont {Wen}},\ }\bibfield  {title} {\bibinfo {title} {Local unitary transformation, long-range quantum entanglement, wave function renormalization, and topological order},\ }\href {https://doi.org/10.1103/PhysRevB.82.155138} {\bibfield  {journal} {\bibinfo  {journal} {Physical Review B}\ }\textbf {\bibinfo {volume} {82}},\ \bibinfo {pages} {155138} (\bibinfo {year} {2010})}\BibitemShut {NoStop}%
\bibitem [{\citenamefont {Harper}\ \emph {et~al.}(2020)\citenamefont {Harper}, \citenamefont {Roy}, \citenamefont {Rudner},\ and\ \citenamefont {Sondhi}}]{HarperTopology2020}%
  \BibitemOpen
  \bibfield  {author} {\bibinfo {author} {\bibfnamefont {F.}~\bibnamefont {Harper}}, \bibinfo {author} {\bibfnamefont {R.}~\bibnamefont {Roy}}, \bibinfo {author} {\bibfnamefont {M.~S.}\ \bibnamefont {Rudner}},\ and\ \bibinfo {author} {\bibfnamefont {S.}~\bibnamefont {Sondhi}},\ }\bibfield  {title} {\bibinfo {title} {Topology and {{Broken Symmetry}} in {{Floquet Systems}}},\ }\href {https://doi.org/10.1146/annurev-conmatphys-031218-013721} {\bibfield  {journal} {\bibinfo  {journal} {Annual Review of Condensed Matter Physics}\ }\textbf {\bibinfo {volume} {11}},\ \bibinfo {pages} {345} (\bibinfo {year} {2020})}\BibitemShut {NoStop}%
\bibitem [{\citenamefont {Eckardt}(2017)}]{EckardtColloquium2017}%
  \BibitemOpen
  \bibfield  {author} {\bibinfo {author} {\bibfnamefont {A.}~\bibnamefont {Eckardt}},\ }\bibfield  {title} {\bibinfo {title} {Colloquium: {{Atomic}} quantum gases in periodically driven optical lattices},\ }\href {https://doi.org/10.1103/RevModPhys.89.011004} {\bibfield  {journal} {\bibinfo  {journal} {Reviews of Modern Physics}\ }\textbf {\bibinfo {volume} {89}},\ \bibinfo {pages} {011004} (\bibinfo {year} {2017})}\BibitemShut {NoStop}%
\bibitem [{\citenamefont {Jotzu}\ \emph {et~al.}(2014)\citenamefont {Jotzu}, \citenamefont {Messer}, \citenamefont {Desbuquois}, \citenamefont {Lebrat}, \citenamefont {Uehlinger}, \citenamefont {Greif},\ and\ \citenamefont {Esslinger}}]{JotzuExperimental2014}%
  \BibitemOpen
  \bibfield  {author} {\bibinfo {author} {\bibfnamefont {G.}~\bibnamefont {Jotzu}}, \bibinfo {author} {\bibfnamefont {M.}~\bibnamefont {Messer}}, \bibinfo {author} {\bibfnamefont {R.}~\bibnamefont {Desbuquois}}, \bibinfo {author} {\bibfnamefont {M.}~\bibnamefont {Lebrat}}, \bibinfo {author} {\bibfnamefont {T.}~\bibnamefont {Uehlinger}}, \bibinfo {author} {\bibfnamefont {D.}~\bibnamefont {Greif}},\ and\ \bibinfo {author} {\bibfnamefont {T.}~\bibnamefont {Esslinger}},\ }\bibfield  {title} {\bibinfo {title} {Experimental realization of the topological {{Haldane}} model with ultracold fermions},\ }\href {https://doi.org/10.1038/nature13915} {\bibfield  {journal} {\bibinfo  {journal} {Nature}\ }\textbf {\bibinfo {volume} {515}},\ \bibinfo {pages} {237} (\bibinfo {year} {2014})}\BibitemShut {NoStop}%
\bibitem [{\citenamefont {Esmann}\ \emph {et~al.}(2018)\citenamefont {Esmann}, \citenamefont {Lamberti}, \citenamefont {Lema{\^i}tre},\ and\ \citenamefont {{Lanzillotti-Kimura}}}]{EsmannTopological2018}%
  \BibitemOpen
  \bibfield  {author} {\bibinfo {author} {\bibfnamefont {M.}~\bibnamefont {Esmann}}, \bibinfo {author} {\bibfnamefont {F.~R.}\ \bibnamefont {Lamberti}}, \bibinfo {author} {\bibfnamefont {A.}~\bibnamefont {Lema{\^i}tre}},\ and\ \bibinfo {author} {\bibfnamefont {N.~D.}\ \bibnamefont {{Lanzillotti-Kimura}}},\ }\bibfield  {title} {\bibinfo {title} {Topological acoustics in coupled nanocavity arrays},\ }\href {https://doi.org/10.1103/PhysRevB.98.161109} {\bibfield  {journal} {\bibinfo  {journal} {Physical Review B}\ }\textbf {\bibinfo {volume} {98}},\ \bibinfo {pages} {161109(R)} (\bibinfo {year} {2018})}\BibitemShut {NoStop}%
\bibitem [{\citenamefont {Yang}\ \emph {et~al.}(2015)\citenamefont {Yang}, \citenamefont {Gao}, \citenamefont {Shi}, \citenamefont {Lin}, \citenamefont {Gao}, \citenamefont {Chong},\ and\ \citenamefont {Zhang}}]{YangTopological2015}%
  \BibitemOpen
  \bibfield  {author} {\bibinfo {author} {\bibfnamefont {Z.}~\bibnamefont {Yang}}, \bibinfo {author} {\bibfnamefont {F.}~\bibnamefont {Gao}}, \bibinfo {author} {\bibfnamefont {X.}~\bibnamefont {Shi}}, \bibinfo {author} {\bibfnamefont {X.}~\bibnamefont {Lin}}, \bibinfo {author} {\bibfnamefont {Z.}~\bibnamefont {Gao}}, \bibinfo {author} {\bibfnamefont {Y.}~\bibnamefont {Chong}},\ and\ \bibinfo {author} {\bibfnamefont {B.}~\bibnamefont {Zhang}},\ }\bibfield  {title} {\bibinfo {title} {Topological {{Acoustics}}},\ }\href {https://doi.org/10.1103/PhysRevLett.114.114301} {\bibfield  {journal} {\bibinfo  {journal} {Physical Review Letters}\ }\textbf {\bibinfo {volume} {114}},\ \bibinfo {pages} {114301} (\bibinfo {year} {2015})}\BibitemShut {NoStop}%
\bibitem [{\citenamefont {Ozawa}\ \emph {et~al.}(2019)\citenamefont {Ozawa}, \citenamefont {Price}, \citenamefont {Amo}, \citenamefont {Goldman}, \citenamefont {Hafezi}, \citenamefont {Lu}, \citenamefont {Rechtsman}, \citenamefont {Schuster}, \citenamefont {Simon}, \citenamefont {Zilberberg},\ and\ \citenamefont {Carusotto}}]{OzawaTopological2019}%
  \BibitemOpen
  \bibfield  {author} {\bibinfo {author} {\bibfnamefont {T.}~\bibnamefont {Ozawa}}, \bibinfo {author} {\bibfnamefont {H.~M.}\ \bibnamefont {Price}}, \bibinfo {author} {\bibfnamefont {A.}~\bibnamefont {Amo}}, \bibinfo {author} {\bibfnamefont {N.}~\bibnamefont {Goldman}}, \bibinfo {author} {\bibfnamefont {M.}~\bibnamefont {Hafezi}}, \bibinfo {author} {\bibfnamefont {L.}~\bibnamefont {Lu}}, \bibinfo {author} {\bibfnamefont {M.~C.}\ \bibnamefont {Rechtsman}}, \bibinfo {author} {\bibfnamefont {D.}~\bibnamefont {Schuster}}, \bibinfo {author} {\bibfnamefont {J.}~\bibnamefont {Simon}}, \bibinfo {author} {\bibfnamefont {O.}~\bibnamefont {Zilberberg}},\ and\ \bibinfo {author} {\bibfnamefont {I.}~\bibnamefont {Carusotto}},\ }\bibfield  {title} {\bibinfo {title} {Topological photonics},\ }\href {https://doi.org/10.1103/RevModPhys.91.015006} {\bibfield  {journal} {\bibinfo  {journal} {Reviews of Modern Physics}\ }\textbf {\bibinfo {volume} {91}},\ \bibinfo {pages} {015006} (\bibinfo {year} {2019})}\BibitemShut {NoStop}%
\bibitem [{\citenamefont {Gross}\ \emph {et~al.}(2012)\citenamefont {Gross}, \citenamefont {Nesme}, \citenamefont {Vogts},\ and\ \citenamefont {Werner}}]{GrossIndex2012}%
  \BibitemOpen
  \bibfield  {author} {\bibinfo {author} {\bibfnamefont {D.}~\bibnamefont {Gross}}, \bibinfo {author} {\bibfnamefont {V.}~\bibnamefont {Nesme}}, \bibinfo {author} {\bibfnamefont {H.}~\bibnamefont {Vogts}},\ and\ \bibinfo {author} {\bibfnamefont {R.~F.}\ \bibnamefont {Werner}},\ }\bibfield  {title} {\bibinfo {title} {Index {{Theory}} of {{One Dimensional Quantum Walks}} and {{Cellular Automata}}},\ }\href {https://doi.org/10.1007/s00220-012-1423-1} {\bibfield  {journal} {\bibinfo  {journal} {Communications in Mathematical Physics}\ }\textbf {\bibinfo {volume} {310}},\ \bibinfo {pages} {419} (\bibinfo {year} {2012})}\BibitemShut {NoStop}%
\bibitem [{\citenamefont {Cedzich}\ \emph {et~al.}(2016)\citenamefont {Cedzich}, \citenamefont {Gr{\"u}nbaum}, \citenamefont {Stahl}, \citenamefont {Vel{\'a}zquez}, \citenamefont {Werner},\ and\ \citenamefont {Werner}}]{CedzichBulkedge2016}%
  \BibitemOpen
  \bibfield  {author} {\bibinfo {author} {\bibfnamefont {C.}~\bibnamefont {Cedzich}}, \bibinfo {author} {\bibfnamefont {F.~A.}\ \bibnamefont {Gr{\"u}nbaum}}, \bibinfo {author} {\bibfnamefont {C.}~\bibnamefont {Stahl}}, \bibinfo {author} {\bibfnamefont {L.}~\bibnamefont {Vel{\'a}zquez}}, \bibinfo {author} {\bibfnamefont {A.~H.}\ \bibnamefont {Werner}},\ and\ \bibinfo {author} {\bibfnamefont {R.~F.}\ \bibnamefont {Werner}},\ }\bibfield  {title} {\bibinfo {title} {Bulk-edge correspondence of one-dimensional quantum walks},\ }\href {https://doi.org/10.1088/1751-8113/49/21/21LT01} {\bibfield  {journal} {\bibinfo  {journal} {Journal of Physics A: Mathematical and Theoretical}\ }\textbf {\bibinfo {volume} {49}},\ \bibinfo {pages} {21LT01} (\bibinfo {year} {2016})}\BibitemShut {NoStop}%
\bibitem [{\citenamefont {Cedzich}\ \emph {et~al.}(2018)\citenamefont {Cedzich}, \citenamefont {Geib}, \citenamefont {Gr{\"u}nbaum}, \citenamefont {Stahl}, \citenamefont {Vel{\'a}zquez}, \citenamefont {Werner},\ and\ \citenamefont {Werner}}]{CedzichTopological2018}%
  \BibitemOpen
  \bibfield  {author} {\bibinfo {author} {\bibfnamefont {C.}~\bibnamefont {Cedzich}}, \bibinfo {author} {\bibfnamefont {T.}~\bibnamefont {Geib}}, \bibinfo {author} {\bibfnamefont {F.~A.}\ \bibnamefont {Gr{\"u}nbaum}}, \bibinfo {author} {\bibfnamefont {C.}~\bibnamefont {Stahl}}, \bibinfo {author} {\bibfnamefont {L.}~\bibnamefont {Vel{\'a}zquez}}, \bibinfo {author} {\bibfnamefont {A.~H.}\ \bibnamefont {Werner}},\ and\ \bibinfo {author} {\bibfnamefont {R.~F.}\ \bibnamefont {Werner}},\ }\bibfield  {title} {\bibinfo {title} {The {{Topological Classification}} of {{One-Dimensional Symmetric Quantum Walks}}},\ }\href {https://doi.org/10.1007/s00023-017-0630-x} {\bibfield  {journal} {\bibinfo  {journal} {Annales Henri Poincar\'e}\ }\textbf {\bibinfo {volume} {19}},\ \bibinfo {pages} {325} (\bibinfo {year} {2018})}\BibitemShut {NoStop}%
\bibitem [{\citenamefont {Cirac}\ \emph {et~al.}(2017)\citenamefont {Cirac}, \citenamefont {{Perez-Garcia}}, \citenamefont {Schuch},\ and\ \citenamefont {Verstraete}}]{CiracMatrix2017}%
  \BibitemOpen
  \bibfield  {author} {\bibinfo {author} {\bibfnamefont {J.~I.}\ \bibnamefont {Cirac}}, \bibinfo {author} {\bibfnamefont {D.}~\bibnamefont {{Perez-Garcia}}}, \bibinfo {author} {\bibfnamefont {N.}~\bibnamefont {Schuch}},\ and\ \bibinfo {author} {\bibfnamefont {F.}~\bibnamefont {Verstraete}},\ }\bibfield  {title} {\bibinfo {title} {Matrix product unitaries: Structure, symmetries, and topological invariants},\ }\href {https://doi.org/10.1088/1742-5468/aa7e55} {\bibfield  {journal} {\bibinfo  {journal} {Journal of Statistical Mechanics: Theory and Experiment}\ }\textbf {\bibinfo {volume} {2017}},\ \bibinfo {pages} {083105} (\bibinfo {year} {2017})}\BibitemShut {NoStop}%
\bibitem [{\citenamefont {Higashikawa}\ \emph {et~al.}(2019)\citenamefont {Higashikawa}, \citenamefont {Nakagawa},\ and\ \citenamefont {Ueda}}]{HigashikawaFloquet2019}%
  \BibitemOpen
  \bibfield  {author} {\bibinfo {author} {\bibfnamefont {S.}~\bibnamefont {Higashikawa}}, \bibinfo {author} {\bibfnamefont {M.}~\bibnamefont {Nakagawa}},\ and\ \bibinfo {author} {\bibfnamefont {M.}~\bibnamefont {Ueda}},\ }\bibfield  {title} {\bibinfo {title} {Floquet chiral magnetic effect},\ }\href {https://doi.org/10.1103/PhysRevLett.123.066403} {\bibfield  {journal} {\bibinfo  {journal} {Physical Review Letters}\ }\textbf {\bibinfo {volume} {123}},\ \bibinfo {pages} {066403} (\bibinfo {year} {2019})}\BibitemShut {NoStop}%
\bibitem [{\citenamefont {Gong}\ \emph {et~al.}(2020)\citenamefont {Gong}, \citenamefont {S{\"u}nderhauf}, \citenamefont {Schuch},\ and\ \citenamefont {Cirac}}]{GongClassification2020}%
  \BibitemOpen
  \bibfield  {author} {\bibinfo {author} {\bibfnamefont {Z.}~\bibnamefont {Gong}}, \bibinfo {author} {\bibfnamefont {C.}~\bibnamefont {S{\"u}nderhauf}}, \bibinfo {author} {\bibfnamefont {N.}~\bibnamefont {Schuch}},\ and\ \bibinfo {author} {\bibfnamefont {J.~I.}\ \bibnamefont {Cirac}},\ }\bibfield  {title} {\bibinfo {title} {Classification of {{Matrix-Product Unitaries}} with {{Symmetries}}},\ }\href {https://doi.org/10.1103/PhysRevLett.124.100402} {\bibfield  {journal} {\bibinfo  {journal} {Physical Review Letters}\ }\textbf {\bibinfo {volume} {124}},\ \bibinfo {pages} {100402} (\bibinfo {year} {2020})}\BibitemShut {NoStop}%
\bibitem [{\citenamefont {Farrelly}(2020)}]{Farrellyreview2020}%
  \BibitemOpen
  \bibfield  {author} {\bibinfo {author} {\bibfnamefont {T.}~\bibnamefont {Farrelly}},\ }\bibfield  {title} {\bibinfo {title} {A review of {{Quantum Cellular Automata}}},\ }\href {https://doi.org/10.22331/q-2020-11-30-368} {\bibfield  {journal} {\bibinfo  {journal} {Quantum}\ }\textbf {\bibinfo {volume} {4}},\ \bibinfo {pages} {368} (\bibinfo {year} {2020})}\BibitemShut {NoStop}%
\bibitem [{\citenamefont {Cedzich}\ \emph {et~al.}(2021)\citenamefont {Cedzich}, \citenamefont {Geib}, \citenamefont {Werner},\ and\ \citenamefont {Werner}}]{CedzichChiral2021}%
  \BibitemOpen
  \bibfield  {author} {\bibinfo {author} {\bibfnamefont {C.}~\bibnamefont {Cedzich}}, \bibinfo {author} {\bibfnamefont {T.}~\bibnamefont {Geib}}, \bibinfo {author} {\bibfnamefont {A.~H.}\ \bibnamefont {Werner}},\ and\ \bibinfo {author} {\bibfnamefont {R.~F.}\ \bibnamefont {Werner}},\ }\bibfield  {title} {\bibinfo {title} {Chiral {{Floquet Systems}} and {{Quantum Walks}} at {{Half-Period}}},\ }\href {https://doi.org/10.1007/s00023-020-00982-6} {\bibfield  {journal} {\bibinfo  {journal} {Annales Henri Poincar\'e}\ }\textbf {\bibinfo {volume} {22}},\ \bibinfo {pages} {375} (\bibinfo {year} {2021})}\BibitemShut {NoStop}%
\bibitem [{\citenamefont {Haah}()}]{HaahTopological2022}%
  \BibitemOpen
  \bibfield  {author} {\bibinfo {author} {\bibfnamefont {J.}~\bibnamefont {Haah}},\ }\bibfield  {title} {\bibinfo {title} {{T}opological phases of unitary dynamics: {C}lassification in {C}lifford category},\ }\bibinfo {note} {\href{https://doi.org/10.48550/arXiv.2205.09141 }{arXiv:2205.09141}}\BibitemShut {NoStop}%
\bibitem [{Note0()}]{Note0}%
  \BibitemOpen
  \bibinfo {note} {The classification in Ref.~\cite {GrossIndex2012} was obtained under the assumption of strict spatial locality, that is, unitaries strictly vanishing outside a finite interval. The classification was later generalized in Ref.~\cite {CedzichChiral2021} to allow for exponential decay.}\BibitemShut {Stop}%
\bibitem [{\citenamefont {Kitaev}(2006)}]{KitaevAnyons2006}%
  \BibitemOpen
  \bibfield  {author} {\bibinfo {author} {\bibfnamefont {A.}~\bibnamefont {Kitaev}},\ }\bibfield  {title} {\bibinfo {title} {Anyons in an exactly solved model and beyond},\ }\href {https://doi.org/10.1016/j.aop.2005.10.005} {\bibfield  {journal} {\bibinfo  {journal} {Annals of Physics}\ }\textbf {\bibinfo {volume} {321}},\ \bibinfo {pages} {2} (\bibinfo {year} {2006})}\BibitemShut {NoStop}%
\bibitem [{\citenamefont {Roy}\ and\ \citenamefont {Harper}(2017)}]{RoyPeriodic2017}%
  \BibitemOpen
  \bibfield  {author} {\bibinfo {author} {\bibfnamefont {R.}~\bibnamefont {Roy}}\ and\ \bibinfo {author} {\bibfnamefont {F.}~\bibnamefont {Harper}},\ }\bibfield  {title} {\bibinfo {title} {Periodic table for {{Floquet}} topological insulators},\ }\href {https://doi.org/10.1103/PhysRevB.96.155118} {\bibfield  {journal} {\bibinfo  {journal} {Physical Review B}\ }\textbf {\bibinfo {volume} {96}},\ \bibinfo {pages} {155118} (\bibinfo {year} {2017})}\BibitemShut {NoStop}%
\bibitem [{\citenamefont {McGinley}\ and\ \citenamefont {Cooper}(2019)}]{McGinleyClassification2019}%
  \BibitemOpen
  \bibfield  {author} {\bibinfo {author} {\bibfnamefont {M.}~\bibnamefont {McGinley}}\ and\ \bibinfo {author} {\bibfnamefont {N.~R.}\ \bibnamefont {Cooper}},\ }\bibfield  {title} {\bibinfo {title} {Classification of topological insulators and superconductors out of equilibrium},\ }\href {https://doi.org/10.1103/PhysRevB.99.075148} {\bibfield  {journal} {\bibinfo  {journal} {Physical Review B}\ }\textbf {\bibinfo {volume} {99}},\ \bibinfo {pages} {075148} (\bibinfo {year} {2019})}\BibitemShut {NoStop}%
\bibitem [{\citenamefont {Liu}(2020)}]{LiuClassification2020}%
  \BibitemOpen
  \bibfield  {author} {\bibinfo {author} {\bibfnamefont {X.}~\bibnamefont {Liu}},\ }\emph {\bibinfo {title} {Classification of {{Static}} and {{Driven Topological}} Insulators}},\ \href {https://escholarship.org/uc/item/4rw7m1kt} {Ph.D. thesis},\ \bibinfo  {school} {UCLA} (\bibinfo {year} {2020})\BibitemShut {NoStop}%
\bibitem [{\citenamefont {Graf}\ and\ \citenamefont {Tauber}(2018)}]{GrafBulk2018}%
  \BibitemOpen
  \bibfield  {author} {\bibinfo {author} {\bibfnamefont {G.~M.}\ \bibnamefont {Graf}}\ and\ \bibinfo {author} {\bibfnamefont {C.}~\bibnamefont {Tauber}},\ }\bibfield  {title} {\bibinfo {title} {Bulk\textendash{{Edge Correspondence}} for {{Two-Dimensional Floquet Topological Insulators}}},\ }\href {https://doi.org/10.1007/s00023-018-0657-7} {\bibfield  {journal} {\bibinfo  {journal} {Annales Henri Poincar\'e}\ }\textbf {\bibinfo {volume} {19}},\ \bibinfo {pages} {709} (\bibinfo {year} {2018})}\BibitemShut {NoStop}%
\bibitem [{\citenamefont {Kitaev}(2009)}]{KitaevPeriodic2009}%
  \BibitemOpen
  \bibfield  {author} {\bibinfo {author} {\bibfnamefont {A.}~\bibnamefont {Kitaev}},\ }\bibfield  {title} {\bibinfo {title} {Periodic table for topological insulators and superconductors},\ }\href {https://doi.org/10.1063/1.3149495} {\bibfield  {journal} {\bibinfo  {journal} {AIP Conference Proceedings}\ }\textbf {\bibinfo {volume} {1134}},\ \bibinfo {pages} {22} (\bibinfo {year} {2009})}\BibitemShut {NoStop}%
\bibitem [{\citenamefont {Altland}\ and\ \citenamefont {Zirnbauer}(1997)}]{AltlandNonstandard1997}%
  \BibitemOpen
  \bibfield  {author} {\bibinfo {author} {\bibfnamefont {A.}~\bibnamefont {Altland}}\ and\ \bibinfo {author} {\bibfnamefont {M.~R.}\ \bibnamefont {Zirnbauer}},\ }\bibfield  {title} {\bibinfo {title} {Nonstandard symmetry classes in mesoscopic normal-superconducting hybrid structures},\ }\href {https://doi.org/10.1103/PhysRevB.55.1142} {\bibfield  {journal} {\bibinfo  {journal} {Physical Review B}\ }\textbf {\bibinfo {volume} {55}},\ \bibinfo {pages} {1142} (\bibinfo {year} {1997})}\BibitemShut {NoStop}%
\bibitem [{\citenamefont {Katsura}\ and\ \citenamefont {Koma}(2018)}]{Katsuranoncommutative2018}%
  \BibitemOpen
  \bibfield  {author} {\bibinfo {author} {\bibfnamefont {H.}~\bibnamefont {Katsura}}\ and\ \bibinfo {author} {\bibfnamefont {T.}~\bibnamefont {Koma}},\ }\bibfield  {title} {\bibinfo {title} {The noncommutative index theorem and the periodic table for disordered topological insulators and superconductors},\ }\href {https://doi.org/10.1063/1.5026964} {\bibfield  {journal} {\bibinfo  {journal} {Journal of Mathematical Physics}\ }\textbf {\bibinfo {volume} {59}},\ \bibinfo {pages} {031903} (\bibinfo {year} {2018})}\BibitemShut {NoStop}%
\bibitem [{\citenamefont {Chung}\ and\ \citenamefont {Shapiro}()}]{ChungTopological2023}%
  \BibitemOpen
  \bibfield  {author} {\bibinfo {author} {\bibfnamefont {J.-H.}\ \bibnamefont {Chung}}\ and\ \bibinfo {author} {\bibfnamefont {J.}~\bibnamefont {Shapiro}},\ }\bibfield  {title} {\bibinfo {title} {{T}opological {C}lassification of {I}nsulators: I. {N}on-interacting {S}pectrally-{G}apped {O}ne-{D}imensional {S}ystems},\ }\bibinfo {note} {\href{https://doi.org/10.48550/arXiv.2306.00268 }{arXiv.2306.00268}}\BibitemShut {NoStop}%
\bibitem [{\citenamefont {Prodan}\ and\ \citenamefont {{Schulz-Baldes}}(2016)}]{ProdanBulk2016}%
  \BibitemOpen
  \bibfield  {author} {\bibinfo {author} {\bibfnamefont {E.}~\bibnamefont {Prodan}}\ and\ \bibinfo {author} {\bibfnamefont {H.}~\bibnamefont {{Schulz-Baldes}}},\ }\href {https://doi.org/10.1007/978-3-319-29351-6} {\emph {\bibinfo {title} {Bulk and {{Boundary Invariants}} for {{Complex Topological Insulators}}}}},\ Mathematical {{Physics Studies}}\ (\bibinfo  {publisher} {{Springer International Publishing}},\ \bibinfo {address} {{Cham}},\ \bibinfo {year} {2016})\BibitemShut {NoStop}%
\bibitem [{\citenamefont {Liu}\ \emph {et~al.}(2018)\citenamefont {Liu}, \citenamefont {Harper},\ and\ \citenamefont {Roy}}]{LiuChiral2018}%
  \BibitemOpen
  \bibfield  {author} {\bibinfo {author} {\bibfnamefont {X.}~\bibnamefont {Liu}}, \bibinfo {author} {\bibfnamefont {F.}~\bibnamefont {Harper}},\ and\ \bibinfo {author} {\bibfnamefont {R.}~\bibnamefont {Roy}},\ }\bibfield  {title} {\bibinfo {title} {Chiral flow in one-dimensional {{Floquet}} topological insulators},\ }\href {https://doi.org/10.1103/PhysRevB.98.165116} {\bibfield  {journal} {\bibinfo  {journal} {Physical Review B}\ }\textbf {\bibinfo {volume} {98}},\ \bibinfo {pages} {165116} (\bibinfo {year} {2018})}\BibitemShut {NoStop}%
\bibitem [{\citenamefont {{Mondragon-Shem}}\ \emph {et~al.}(2014)\citenamefont {{Mondragon-Shem}}, \citenamefont {Hughes}, \citenamefont {Song},\ and\ \citenamefont {Prodan}}]{Mondragon-ShemTopological2014}%
  \BibitemOpen
  \bibfield  {author} {\bibinfo {author} {\bibfnamefont {I.}~\bibnamefont {{Mondragon-Shem}}}, \bibinfo {author} {\bibfnamefont {T.~L.}\ \bibnamefont {Hughes}}, \bibinfo {author} {\bibfnamefont {J.}~\bibnamefont {Song}},\ and\ \bibinfo {author} {\bibfnamefont {E.}~\bibnamefont {Prodan}},\ }\bibfield  {title} {\bibinfo {title} {Topological {{Criticality}} in the {{Chiral-Symmetric AIII Class}} at {{Strong Disorder}}},\ }\href {https://doi.org/10.1103/PhysRevLett.113.046802} {\bibfield  {journal} {\bibinfo  {journal} {Physical Review Letters}\ }\textbf {\bibinfo {volume} {113}},\ \bibinfo {pages} {046802} (\bibinfo {year} {2014})}\BibitemShut {NoStop}%
\bibitem [{\citenamefont {Teo}\ and\ \citenamefont {Kane}(2010)}]{TeoTopological2010}%
  \BibitemOpen
  \bibfield  {author} {\bibinfo {author} {\bibfnamefont {J.~C.~Y.}\ \bibnamefont {Teo}}\ and\ \bibinfo {author} {\bibfnamefont {C.~L.}\ \bibnamefont {Kane}},\ }\bibfield  {title} {\bibinfo {title} {Topological defects and gapless modes in insulators and superconductors},\ }\href {https://doi.org/10.1103/PhysRevB.82.115120} {\bibfield  {journal} {\bibinfo  {journal} {Physical Review B}\ }\textbf {\bibinfo {volume} {82}},\ \bibinfo {pages} {115120} (\bibinfo {year} {2010})}\BibitemShut {NoStop}%
\bibitem [{\citenamefont {Kitaev}(2001)}]{KitaevUnpaired2001}%
  \BibitemOpen
  \bibfield  {author} {\bibinfo {author} {\bibfnamefont {A.~Y.}\ \bibnamefont {Kitaev}},\ }\bibfield  {title} {\bibinfo {title} {Unpaired {{Majorana}} fermions in quantum wires},\ }\href {https://doi.org/10.1070/1063-7869/44/10S/S29} {\bibfield  {journal} {\bibinfo  {journal} {Physics-Uspekhi}\ }\textbf {\bibinfo {volume} {44}},\ \bibinfo {pages} {131} (\bibinfo {year} {2001})}\BibitemShut {NoStop}%
\bibitem [{\citenamefont {{B. Bernevig with T. Hughes}}(2013)}]{BernevigTopological2013}%
  \BibitemOpen
  \bibfield  {author} {\bibinfo {author} {\bibnamefont {{B. Bernevig with T. Hughes}}},\ }\href {https://doi.org/10.1515/9781400846733} {\emph {\bibinfo {title} {Topological Insulators and Topological Superconductors}}}\ (\bibinfo  {publisher} {{Princeton University Press}},\ \bibinfo {address} {Princeton and Oxford},\ \bibinfo {year} {2013})\BibitemShut {NoStop}%
\bibitem [{\citenamefont {Roy}()}]{RoyInteger2006}%
  \BibitemOpen
  \bibfield  {author} {\bibinfo {author} {\bibfnamefont {R.}~\bibnamefont {Roy}},\ }\bibfield  {title} {\bibinfo {title} {{I}nteger {Q}uantum {H}all {E}ffect on a {S}quare {L}attice with {Z}ero {N}et {M}agnetic {F}ield},\ }\bibinfo {note} {\href{https://doi.org/10.48550/arXiv.cond-mat/0603271 }{arXiv:cond-mat/0603271}}\BibitemShut {NoStop}%
\bibitem [{Note1()}]{Note1}%
  \BibitemOpen
  \bibinfo {note} {Here we have corrected typos in Eq. (1) of Ref.~\cite {RoyInteger2006}: the second left parentheses in the first line and the last right parentheses in the second line should be deleted.}\BibitemShut {Stop}%
\bibitem [{\citenamefont {Haah}\ \emph {et~al.}(2023)\citenamefont {Haah}, \citenamefont {Fidkowski},\ and\ \citenamefont {Hastings}}]{HaahNontrivial2023}%
  \BibitemOpen
  \bibfield  {author} {\bibinfo {author} {\bibfnamefont {J.}~\bibnamefont {Haah}}, \bibinfo {author} {\bibfnamefont {L.}~\bibnamefont {Fidkowski}},\ and\ \bibinfo {author} {\bibfnamefont {M.~B.}\ \bibnamefont {Hastings}},\ }\bibfield  {title} {\bibinfo {title} {Nontrivial {{Quantum Cellular Automata}} in {{Higher Dimensions}}},\ }\href {https://doi.org/10.1007/s00220-022-04528-1} {\bibfield  {journal} {\bibinfo  {journal} {Communications in Mathematical Physics}\ }\textbf {\bibinfo {volume} {398}},\ \bibinfo {pages} {469} (\bibinfo {year} {2023})}\BibitemShut {NoStop}%
\bibitem [{\citenamefont {Geiko}\ and\ \citenamefont {Hu}()}]{GeikoHomotopy2023}%
  \BibitemOpen
  \bibfield  {author} {\bibinfo {author} {\bibfnamefont {R.}~\bibnamefont {Geiko}}\ and\ \bibinfo {author} {\bibfnamefont {Y.}~\bibnamefont {Hu}},\ }\bibfield  {title} {\bibinfo {title} {{H}omotopy {C}lassification of loops of {C}lifford unitaries},\ }\bibinfo {note} {\href{ https://doi.org/10.48550/arXiv.2306.09903}{arXiv.2306.09903}}\BibitemShut {NoStop}%
\end{thebibliography}%

\end{document}